\RequirePackage{snapshot}
\documentclass[longauth]{aa}
\usepackage{amsmath,amssymb,amsfonts}
\usepackage{graphicx}
\usepackage{txfonts}
\usepackage{enumerate} 
\usepackage{colordvi} 
\usepackage{bm}
\usepackage{epsfig}
\usepackage{hyperref}
\usepackage[dvipsnames]{xcolor}
\usepackage{color, colortbl}
\usepackage[utf8]{inputenc} 
\usepackage{braket}
\usepackage{euclid}
\usepackage{placeins}
\usepackage{tabularx}
\usepackage{booktabs}
\usepackage{diagbox}
\usepackage{adjustbox}
\usepackage{comment}
\usepackage{multirow}
\bibpunct[, ]{(}{)}{,}{a}{,}{,}

\let\laplacian\Delta
\let\angle\Omega
\let\Gamma\varGamma
\let\Delta\varDelta
\let\Theta\varTheta
\let\Xi\varXi
\let\Pi\varPi
\let\Sigma\varSigma
\let\Upsilon\varUpsilon
\let\Phi\varPhi
\let\Psi\varPsi
\let\Omega\varOmega
\newcommand{\numberdensity}{\ensuremath{\mathcal{N}}}

\newcommand{\sigmastarsquared}{\ensuremath{\sigma_8^2(z_*)}}

\def\fra#1{{#1}} 

\definecolor{amaranth}{rgb}{0.9, 0.17, 0.31}
\definecolor{forestgreen(web)}{rgb}{0.13, 0.55, 0.13}
\definecolor{lavender(web)}{rgb}{0.9, 0.9, 0.98}
\definecolor{cosmiclatte}{rgb}{1.0, 0.97, 0.91}
\definecolor{jonquil}{rgb}{0.98, 0.85, 0.37}
\definecolor{khaki(x11)(lightkhaki)}{rgb}{0.94, 0.9, 0.55}
\definecolor{thistle}{rgb}{0.85, 0.75, 0.85}

\newcommand{\coffe}{\texttt{coffe}}
\newcommand{\coffeurl}{https://github.com/JCGoran/coffe}
\newcommand{\fitk}{\texttt{FITK}}
\newcommand{\fitkurl}{https://github.com/JCGoran/fitk}
\newcommand{\cbl}{\texttt{CosmoBolognaLib}}
\newcommand{\cblurl}{https://github.com/federicomarulli/CosmoBolognaLib}

\newcommand{\lcdm}{$\Lambda$CDM}
\newcommand{\wcdm}{$\wzero\wa$CDM}
\newcommand{\wzero}{w_0}
\newcommand{\wa}{w_a}
\newcommand{\Omm}{\Omega_{{\rm m},0}}
\newcommand{\Omb}{\Omega_{{\rm b},0}}
\newcommand{\omm}{\omega_{{\rm m},0}}
\newcommand{\omb}{\omega_{{\rm b},0}}

\renewcommand{\bea}{\begin{eqnarray}}
\renewcommand{\eea}{\end{eqnarray}}
\newcommand{\bn}{{{\bm n}}}
\newcommand{\bV}{{\bm V}}
\newcommand{\De}{\Delta}
\newcommand{\HH}{\mathcal{H}}

\newcommand{\de}{\mathrm{d}}

\renewcommand{\la}{\lambda}
\newcommand{\Om}{\Omega}

\newcommand{\ns}{n_{\text{s}}}

\newcommand{\rsource}{r_\text{s}}

\newcommand{\fisher}{\tens{F}}
\newcommand{\covariance}{\tens{C}}

\newcommand{\std}{\mathrm{std}}
\newcommand{\magn}{\ensuremath{\text{magn}}}
\newcommand{\dd}{\partial}

\newcommand{\minus}{\ensuremath{-}}

\defcitealias{Blanchard:2019oqi}{EP:VII}
\defcitealias{2016A&A...594A..14P}{Planck Collaboration, 2016}

\usepackage[nameinlink,noabbrev]{cleveref}
\Crefname{equation}{Eq.}{Eqs}
\Crefname{section}{Sect.}{Sects}
\Crefname{figure}{Fig.}{Figs}
\crefname{equation}{Equation}{Equations}
\crefname{section}{Section}{Sections}
\crefname{figure}{Figure}{Figures}
\usepackage[switch, modulo]{lineno}

\begin{document}
\title{\Euclid preparation}
\subtitle{TBD. Impact of magnification on spectroscopic galaxy clustering}

\providecommand{\orcid}[1]{} 
\renewcommand{\orcid}[1]{}
\author{Euclid Collaboration: G.~Jelic-Cizmek\orcid{0000-0002-2773-1765}$^{1}$\thanks{\email{goran.jelic-cizmek@unige.ch}}, F.~Sorrenti\orcid{0000-0001-7141-9659}$^{1}$, F.~Lepori$^{2}$, C.~Bonvin\orcid{0000-0002-5318-4064}$^{1}$, S.~Camera\orcid{0000-0003-3399-3574}$^{3,4,5}$, F.~J.~Castander\orcid{0000-0001-7316-4573}$^{6,7}$, R.~Durrer\orcid{0000-0001-9833-2086}$^{1}$, P.~Fosalba\orcid{0000-0002-1510-5214}$^{7,8}$, M.~Kunz\orcid{0000-0002-3052-7394}$^{1}$, L.~Lombriser$^{1}$
% ...
}

\institute{$^{1}$ Universit\'e de Gen\`eve, D\'epartement de Physique Th\'eorique and Centre for Astroparticle Physics, 24 quai Ernest-Ansermet, CH-1211 Gen\`eve 4, Switzerland\\
$^{2}$ Institute for Computational Science, University of Zurich, Winterthurerstrasse 190, 8057 Zurich, Switzerland\\
$^{3}$ Dipartimento di Fisica, Universit\`a degli Studi di Torino, Via P. Giuria 1, 10125 Torino, Italy\\
$^{4}$ INFN-Sezione di Torino, Via P. Giuria 1, 10125 Torino, Italy\\
$^{5}$ INAF-Osservatorio Astrofisico di Torino, Via Osservatorio 20, 10025 Pino Torinese (TO), Italy\\
$^{6}$ Institute of Space Sciences (ICE, CSIC), Campus UAB, Carrer de Can Magrans, s/n, 08193 Barcelona, Spain\\
$^{7}$ Institut d'Estudis Espacials de Catalunya (IEEC), Carrer Gran Capit\'a 2-4, 08034 Barcelona, Spain\\
$^{8}$ Institut de Ciencies de l'Espai (IEEC-CSIC), Campus UAB, Carrer de Can Magrans, s/n Cerdanyola del Vall\'es, 08193 Barcelona, Spain\\
$^{9}$ Institut de Recherche en Astrophysique et Plan\'etologie (IRAP), Universit\'e de Toulouse, CNRS, UPS, CNES, 14 Av. Edouard Belin, 31400 Toulouse, France\\
$^{10}$ Institut f\"ur Theoretische Physik, University of Heidelberg, Philosophenweg 16, 69120 Heidelberg, Germany\\
$^{11}$ Universit\'e St Joseph; Faculty of Sciences, Beirut, Lebanon\\
$^{12}$ Universit\'e Paris-Saclay, CNRS, Institut d'astrophysique spatiale, 91405, Orsay, France\\
$^{13}$ Institute of Cosmology and Gravitation, University of Portsmouth, Portsmouth PO1 3FX, UK\\
$^{14}$ INAF-Osservatorio Astronomico di Brera, Via Brera 28, 20122 Milano, Italy\\
$^{15}$ Dipartimento di Fisica e Astronomia, Universit\`a di Bologna, Via Gobetti 93/2, 40129 Bologna, Italy\\
$^{16}$ INAF-Osservatorio di Astrofisica e Scienza dello Spazio di Bologna, Via Piero Gobetti 93/3, 40129 Bologna, Italy\\
$^{17}$ INFN-Sezione di Bologna, Viale Berti Pichat 6/2, 40127 Bologna, Italy\\
$^{18}$ Max Planck Institute for Extraterrestrial Physics, Giessenbachstr. 1, 85748 Garching, Germany\\
$^{19}$ Dipartimento di Fisica, Universit\`a di Genova, Via Dodecaneso 33, 16146, Genova, Italy\\
$^{20}$ INFN-Sezione di Genova, Via Dodecaneso 33, 16146, Genova, Italy\\
$^{21}$ Department of Physics "E. Pancini", University Federico II, Via Cinthia 6, 80126, Napoli, Italy\\
$^{22}$ INAF-Osservatorio Astronomico di Capodimonte, Via Moiariello 16, 80131 Napoli, Italy\\
$^{23}$ INFN section of Naples, Via Cinthia 6, 80126, Napoli, Italy\\
$^{24}$ Instituto de Astrof\'isica e Ci\^encias do Espa\c{c}o, Universidade do Porto, CAUP, Rua das Estrelas, PT4150-762 Porto, Portugal\\
$^{25}$ INAF-IASF Milano, Via Alfonso Corti 12, 20133 Milano, Italy\\
$^{26}$ INAF-Osservatorio Astronomico di Roma, Via Frascati 33, 00078 Monteporzio Catone, Italy\\
$^{27}$ INFN-Sezione di Roma, Piazzale Aldo Moro, 2 - c/o Dipartimento di Fisica, Edificio G. Marconi, 00185 Roma, Italy\\
$^{28}$ Institut de F\'{i}sica d'Altes Energies (IFAE), The Barcelona Institute of Science and Technology, Campus UAB, 08193 Bellaterra (Barcelona), Spain\\
$^{29}$ Port d'Informaci\'{o} Cient\'{i}fica, Campus UAB, C. Albareda s/n, 08193 Bellaterra (Barcelona), Spain\\
$^{30}$ Institute for Theoretical Particle Physics and Cosmology (TTK), RWTH Aachen University, 52056 Aachen, Germany\\
$^{31}$ Dipartimento di Fisica e Astronomia "Augusto Righi" - Alma Mater Studiorum Universit\`a di Bologna, Viale Berti Pichat 6/2, 40127 Bologna, Italy\\
$^{32}$ Institute for Astronomy, University of Edinburgh, Royal Observatory, Blackford Hill, Edinburgh EH9 3HJ, UK\\
$^{33}$ Jodrell Bank Centre for Astrophysics, Department of Physics and Astronomy, University of Manchester, Oxford Road, Manchester M13 9PL, UK\\
$^{34}$ European Space Agency/ESRIN, Largo Galileo Galilei 1, 00044 Frascati, Roma, Italy\\
$^{35}$ ESAC/ESA, Camino Bajo del Castillo, s/n., Urb. Villafranca del Castillo, 28692 Villanueva de la Ca\~nada, Madrid, Spain\\
$^{36}$ University of Lyon, Univ Claude Bernard Lyon 1, CNRS/IN2P3, IP2I Lyon, UMR 5822, 69622 Villeurbanne, France\\
$^{37}$ Institute of Physics, Laboratory of Astrophysics, Ecole Polytechnique F\'ed\'erale de Lausanne (EPFL), Observatoire de Sauverny, 1290 Versoix, Switzerland\\
$^{38}$ UCB Lyon 1, CNRS/IN2P3, IUF, IP2I Lyon, 4 rue Enrico Fermi, 69622 Villeurbanne, France\\
$^{39}$ Mullard Space Science Laboratory, University College London, Holmbury St Mary, Dorking, Surrey RH5 6NT, UK\\
$^{40}$ Department of Astronomy, University of Geneva, ch. d'Ecogia 16, 1290 Versoix, Switzerland\\
$^{41}$ INAF-Istituto di Astrofisica e Planetologia Spaziali, via del Fosso del Cavaliere, 100, 00100 Roma, Italy\\
$^{42}$ Instituto de Astrof\'isica e Ci\^encias do Espa\c{c}o, Faculdade de Ci\^encias, Universidade de Lisboa, Campo Grande, 1749-016 Lisboa, Portugal\\
$^{43}$ Departamento de F\'isica, Faculdade de Ci\^encias, Universidade de Lisboa, Edif\'icio C8, Campo Grande, PT1749-016 Lisboa, Portugal\\
$^{44}$ INFN-Padova, Via Marzolo 8, 35131 Padova, Italy\\
$^{45}$ Universit\'e Paris-Saclay, Universit\'e Paris Cit\'e, CEA, CNRS, AIM, 91191, Gif-sur-Yvette, France\\
$^{46}$ INAF-Osservatorio Astronomico di Trieste, Via G. B. Tiepolo 11, 34143 Trieste, Italy\\
$^{47}$ Istituto Nazionale di Fisica Nucleare, Sezione di Bologna, Via Irnerio 46, 40126 Bologna, Italy\\
$^{48}$ INAF-Osservatorio Astronomico di Padova, Via dell'Osservatorio 5, 35122 Padova, Italy\\
$^{49}$ University Observatory, Faculty of Physics, Ludwig-Maximilians-Universit{\"a}t, Scheinerstr. 1, 81679 Munich, Germany\\
$^{50}$ Institute of Theoretical Astrophysics, University of Oslo, P.O. Box 1029 Blindern, 0315 Oslo, Norway\\
$^{51}$ Leiden Observatory, Leiden University, Niels Bohrweg 2, 2333 CA Leiden, The Netherlands\\
$^{52}$ Jet Propulsion Laboratory, California Institute of Technology, 4800 Oak Grove Drive, Pasadena, CA, 91109, USA\\
$^{53}$ von Hoerner \& Sulger GmbH, Schlo{\ss}Platz 8, 68723 Schwetzingen, Germany\\
$^{54}$ Technical University of Denmark, Elektrovej 327, 2800 Kgs. Lyngby, Denmark\\
$^{55}$ Cosmic Dawn Center (DAWN), Denmark\\
$^{56}$ Max-Planck-Institut f\"ur Astronomie, K\"onigstuhl 17, 69117 Heidelberg, Germany\\
$^{57}$ Department of Physics and Helsinki Institute of Physics, Gustaf H\"allstr\"omin katu 2, 00014 University of Helsinki, Finland\\
$^{58}$ Aix-Marseille Universit\'e, CNRS/IN2P3, CPPM, Marseille, France\\
$^{59}$ AIM, CEA, CNRS, Universit\'{e} Paris-Saclay, Universit\'{e} de Paris, 91191 Gif-sur-Yvette, France\\
$^{60}$ Department of Physics, P.O. Box 64, 00014 University of Helsinki, Finland\\
$^{61}$ Helsinki Institute of Physics, Gustaf H{\"a}llstr{\"o}min katu 2, University of Helsinki, Helsinki, Finland\\
$^{62}$ NOVA optical infrared instrumentation group at ASTRON, Oude Hoogeveensedijk 4, 7991PD, Dwingeloo, The Netherlands\\
$^{63}$ Universit\"at Bonn, Argelander-Institut f\"ur Astronomie, Auf dem H\"ugel 71, 53121 Bonn, Germany\\
$^{64}$ Aix-Marseille Universit\'e, CNRS, CNES, LAM, Marseille, France\\
$^{65}$ Dipartimento di Fisica e Astronomia "Augusto Righi" - Alma Mater Studiorum Universit\`a di Bologna, via Piero Gobetti 93/2, 40129 Bologna, Italy\\
$^{66}$ Department of Physics, Institute for Computational Cosmology, Durham University, South Road, DH1 3LE, UK\\
$^{67}$ Universit\'e Paris Cit\'e, CNRS, Astroparticule et Cosmologie, 75013 Paris, France\\
$^{68}$ European Space Agency/ESTEC, Keplerlaan 1, 2201 AZ Noordwijk, The Netherlands\\
$^{69}$ Department of Physics and Astronomy, University of Aarhus, Ny Munkegade 120, DK-8000 Aarhus C, Denmark\\
$^{70}$ Centre for Astrophysics, University of Waterloo, Waterloo, Ontario N2L 3G1, Canada\\
$^{71}$ Department of Physics and Astronomy, University of Waterloo, Waterloo, Ontario N2L 3G1, Canada\\
$^{72}$ Perimeter Institute for Theoretical Physics, Waterloo, Ontario N2L 2Y5, Canada\\
$^{73}$ Universit\'e Paris-Saclay, Universit\'e Paris Cit\'e, CEA, CNRS, Astrophysique, Instrumentation et Mod\'elisation Paris-Saclay, 91191 Gif-sur-Yvette, France\\
$^{74}$ Space Science Data Center, Italian Space Agency, via del Politecnico snc, 00133 Roma, Italy\\
$^{75}$ Centre National d'Etudes Spatiales -- Centre spatial de Toulouse, 18 avenue Edouard Belin, 31401 Toulouse Cedex 9, France\\
$^{76}$ Institute of Space Science, Str. Atomistilor, nr. 409 M\u{a}gurele, Ilfov, 077125, Romania\\
$^{77}$ Instituto de Astrof\'isica de Canarias, Calle V\'ia L\'actea s/n, 38204, San Crist\'obal de La Laguna, Tenerife, Spain\\
$^{78}$ Departamento de Astrof\'isica, Universidad de La Laguna, 38206, La Laguna, Tenerife, Spain\\
$^{79}$ Dipartimento di Fisica e Astronomia "G. Galilei", Universit\`a di Padova, Via Marzolo 8, 35131 Padova, Italy\\
$^{80}$ Universit\"ats-Sternwarte M\"unchen, Fakult\"at f\"ur Physik, Ludwig-Maximilians-Universit\"at M\"unchen, Scheinerstrasse 1, 81679 M\"unchen, Germany\\
$^{81}$ Departamento de F\'isica, FCFM, Universidad de Chile, Blanco Encalada 2008, Santiago, Chile\\
$^{82}$ Universit\"at Innsbruck, Institut f\"ur Astro- und Teilchenphysik, Technikerstr. 25/8, 6020 Innsbruck, Austria\\
$^{83}$ Satlantis, University Science Park, Sede Bld 48940, Leioa-Bilbao, Spain\\
$^{84}$ Centro de Investigaciones Energ\'eticas, Medioambientales y Tecnol\'ogicas (CIEMAT), Avenida Complutense 40, 28040 Madrid, Spain\\
$^{85}$ Instituto de Astrof\'isica e Ci\^encias do Espa\c{c}o, Faculdade de Ci\^encias, Universidade de Lisboa, Tapada da Ajuda, 1349-018 Lisboa, Portugal\\
$^{86}$ Universidad Polit\'ecnica de Cartagena, Departamento de Electr\'onica y Tecnolog\'ia de Computadoras,  Plaza del Hospital 1, 30202 Cartagena, Spain\\
$^{87}$ Kapteyn Astronomical Institute, University of Groningen, PO Box 800, 9700 AV Groningen, The Netherlands\\
$^{88}$ INFN-Bologna, Via Irnerio 46, 40126 Bologna, Italy\\
$^{89}$ Infrared Processing and Analysis Center, California Institute of Technology, Pasadena, CA 91125, USA\\
$^{90}$ IFPU, Institute for Fundamental Physics of the Universe, via Beirut 2, 34151 Trieste, Italy\\
$^{91}$ Institut d'Astrophysique de Paris, 98bis Boulevard Arago, 75014, Paris, France\\
$^{92}$ Junia, EPA department, 41 Bd Vauban, 59800 Lille, France\\
$^{93}$ SISSA, International School for Advanced Studies, Via Bonomea 265, 34136 Trieste TS, Italy\\
$^{94}$ INFN, Sezione di Trieste, Via Valerio 2, 34127 Trieste TS, Italy\\
$^{95}$ ICSC - Centro Nazionale di Ricerca in High Performance Computing, Big Data e Quantum Computing, Via Magnanelli 2, Bologna, Italy\\
$^{96}$ Instituto de F\'isica Te\'orica UAM-CSIC, Campus de Cantoblanco, 28049 Madrid, Spain\\
$^{97}$ CERCA/ISO, Department of Physics, Case Western Reserve University, 10900 Euclid Avenue, Cleveland, OH 44106, USA\\
$^{98}$ Laboratoire Univers et Th\'eorie, Observatoire de Paris, Universit\'e PSL, Universit\'e Paris Cit\'e, CNRS, 92190 Meudon, France\\
$^{99}$ Dipartimento di Fisica e Scienze della Terra, Universit\`a degli Studi di Ferrara, Via Giuseppe Saragat 1, 44122 Ferrara, Italy\\
$^{100}$ Istituto Nazionale di Fisica Nucleare, Sezione di Ferrara, Via Giuseppe Saragat 1, 44122 Ferrara, Italy\\
$^{101}$ Minnesota Institute for Astrophysics, University of Minnesota, 116 Church St SE, Minneapolis, MN 55455, USA\\
$^{102}$ INAF, Istituto di Radioastronomia, Via Piero Gobetti 101, 40129 Bologna, Italy\\
$^{103}$ Universit\'e C\^{o}te d'Azur, Observatoire de la C\^{o}te d'Azur, CNRS, Laboratoire Lagrange, Bd de l'Observatoire, CS 34229, 06304 Nice cedex 4, France\\
$^{104}$ Institute Lorentz, Leiden University, PO Box 9506, Leiden 2300 RA, The Netherlands\\
$^{105}$ Institute for Astronomy, University of Hawaii, 2680 Woodlawn Drive, Honolulu, HI 96822, USA\\
$^{106}$ Department of Physics \& Astronomy, University of California Irvine, Irvine CA 92697, USA\\
$^{107}$ Department of Astronomy \& Physics and Institute for Computational Astrophysics, Saint Mary's University, 923 Robie Street, Halifax, Nova Scotia, B3H 3C3, Canada\\
$^{108}$ Departamento F\'isica Aplicada, Universidad Polit\'ecnica de Cartagena, Campus Muralla del Mar, 30202 Cartagena, Murcia, Spain\\
$^{109}$ Department of Physics, Oxford University, Keble Road, Oxford OX1 3RH, UK\\
$^{110}$ Department of Computer Science, Aalto University, PO Box 15400, Espoo, FI-00 076, Finland\\
$^{111}$ Ruhr University Bochum, Faculty of Physics and Astronomy, Astronomical Institute (AIRUB), German Centre for Cosmological Lensing (GCCL), 44780 Bochum, Germany\\
$^{112}$ Universit\'e Paris-Saclay, CNRS/IN2P3, IJCLab, 91405 Orsay, France\\
$^{113}$ Univ. Grenoble Alpes, CNRS, Grenoble INP, LPSC-IN2P3, 53, Avenue des Martyrs, 38000, Grenoble, France\\
$^{114}$ Department of Physics and Astronomy, Vesilinnantie 5, 20014 University of Turku, Finland\\
$^{115}$ Serco for European Space Agency (ESA), Camino bajo del Castillo, s/n, Urbanizacion Villafranca del Castillo, Villanueva de la Ca\~nada, 28692 Madrid, Spain\\
$^{116}$ ARC Centre of Excellence for Dark Matter Particle Physics, Melbourne, Australia\\
$^{117}$ Centre for Astrophysics \& Supercomputing, Swinburne University of Technology, Victoria 3122, Australia\\
$^{118}$ Oskar Klein Centre for Cosmoparticle Physics, Department of Physics, Stockholm University, Stockholm, SE-106 91, Sweden\\
$^{119}$ Astrophysics Group, Blackett Laboratory, Imperial College London, London SW7 2AZ, UK\\
$^{120}$ Centre de Calcul de l'IN2P3/CNRS, 21 avenue Pierre de Coubertin 69627 Villeurbanne Cedex, France\\
$^{121}$ Dipartimento di Fisica, Sapienza Universit\`a di Roma, Piazzale Aldo Moro 2, 00185 Roma, Italy\\
$^{122}$ Centro de Astrof\'{\i}sica da Universidade do Porto, Rua das Estrelas, 4150-762 Porto, Portugal\\
$^{123}$ Zentrum f\"ur Astronomie, Universit\"at Heidelberg, Philosophenweg 12, 69120 Heidelberg, Germany\\
$^{124}$ Dipartimento di Fisica, Universit\`a di Roma Tor Vergata, Via della Ricerca Scientifica 1, Roma, Italy\\
$^{125}$ INFN, Sezione di Roma 2, Via della Ricerca Scientifica 1, Roma, Italy\\
$^{126}$ Dipartimento di Fisica - Sezione di Astronomia, Universit\`a di Trieste, Via Tiepolo 11, 34131 Trieste, Italy\\
$^{127}$ Department of Astrophysical Sciences, Peyton Hall, Princeton University, Princeton, NJ 08544, USA\\
$^{128}$ Niels Bohr Institute, University of Copenhagen, Jagtvej 128, 2200 Copenhagen, Denmark\\
$^{129}$ Cosmic Dawn Center (DAWN)}

\date{\today}

\authorrunning{WP9 et al.}

\titlerunning{\Euclid preparation: Impact of magnification on spectroscopic galaxy clustering}

\abstract{
In this paper we investigate the impact of lensing magnification on the analysis of \Euclid's spectroscopic survey, using the multipoles of the 2-point correlation function {for galaxy clustering}. We determine the impact of lensing magnification on cosmological constraints, and the {expected} shift in the best-fit parameters {if magnification is ignored}. We consider two cosmological analyses: i) a full-shape analysis based on the $\Lambda$CDM model and its extension $\wzero\wa$CDM and ii) a model-independent analysis that measures the growth rate of structure in each redshift bin.
We adopt two complementary approaches in our forecast: the Fisher matrix formalism and the Markov chain Monte Carlo method.
The fiducial values of the local count slope (or magnification bias), which regulates the amplitude of the lensing magnification, have been estimated from the \Euclid Flagship simulations. We use linear perturbation theory and model the 2-point correlation function with the public code \coffe{}.
For a $\Lambda$CDM model, we find that the estimation of cosmological parameters is biased at the level of 0.4--0.7 standard deviations, while for a $\wzero\wa$CDM dynamical dark energy model, lensing magnification has a somewhat smaller impact, with shifts below 0.5 standard deviations.
In a model-independent analysis aiming to measure the growth rate of structure, we find that the estimation of the growth rate is biased by up to $1.2$ standard deviations in the highest redshift bin.
As a result, lensing magnification cannot be neglected in the spectroscopic survey, especially if we want to determine the growth factor, one of the most promising ways to test general relativity with \Euclid.
We also find that, by including lensing magnification with a simple template, this shift {can be almost entirely eliminated with minimal computational overhead}.
}

   \keywords{Cosmology -- large-scale structure of Universe -- cosmological parameters -- Cosmology: theory}

   \maketitle

\section{Introduction}
The European Space Agency's \Euclid satellite mission \citep{Laureijs:2011gra,Amendola:2016saw} aims at shedding new light on the so-called dark components of the Universe, namely dark matter and dark energy. {Dark matter}, a mysterious form of matter that does not seem to emit light, yet it accounts for more than $80\%$ of the total matter content of the Universe, forms the bulk of the large-scale cosmic structure, upon which galaxies form and evolve. The latter is even more elusive, and {it} is what drives the current accelerated expansion of the Universe, contributing to about $70\%$ of the total cosmic energy budget~\citep[see, e.g.,][for a review of the current concordance cosmological model and the main theoretical challenges it faces]{2016PDU....12...56B}.
In fact, there is another possibility to explain the effects we ascribe to (either or both) dark matter and dark energy: that the theory we use to analyse the data is incorrect. This approach goes under the name of `modified gravity' \citep[e.g.,][]{2012PhR...513....1C}. Thanks to the extent and exquisite precision of \Euclid's data, we shall soon be able to further test general relativity on scales far from the strong-gravity regime where it has been tested to supreme precision~\citep[see, e.g.,][for a review of the current status]{Cardoso:2019rvt}.

\Euclid will consist of two primary probes: a catalogue of about 30 million galaxies with spectroscopic redshift information, spanning a redshift range between $z = 0.8$ and $z = 1.8$, and a catalogue of 1.5 billion galaxy images with photometric redshifts down to $z = 2$~\citep[see][for further details on the specifics of \Euclid surveys]{Laureijs:2011gra, Amendola:2016saw}.
One of the main goals of the spectroscopic survey of \Euclid is to measure the so-called growth rate which is very sensitive to the theory of gravity, see for example~\citet{Alam:2016hwk}.
However, in order to robustly test alternatives to general relativity, it is crucial to take into account in the analysis all of the relevant effects. One effect that has been overlooked in previous forecasts regarding the performance of the \Euclid spectroscopic survey is lensing magnification~\citep{Matsubara:2004fr}.

The aim of this paper is to investigate whether lensing magnification has to be included in this analysis.
It is well known that lensing magnification has to be included in a photometric survey for correct estimation of cosmological parameters{~\citep{Duncan:2013haa, Cardona:2016qxn, Villa_2018, Lorenz:2017iez, Jelic-Cizmek:2020pkh, Euclid:2021rez, LSSTDarkEnergyScience:2021bah}}. However, as the density and redshift space distortion contributions are significantly larger in a spectroscopic survey, one might hope that lensing magnification can be neglected in this case. In this paper, we show that this is not the case, and that neglecting lensing can shift the inferred cosmological parameters by up to $0.7\sigma$, and can affect the measured growth rate by up to $1\sigma$.
We then propose a method to reduce the shifts to an acceptable level. This method consists in adding the lensing magnification signal into the modelling, using fixed cosmological parameters in $\Lambda$CDM. Of course, in this way the lensing magnification is not exactly correct (since we do not know the theory of gravity nor the cosmological parameters), but we show that this is enough to de-bias the analysis, reducing the shifts to less than $0.1\sigma$.

The paper is structured as follows. In the next section, we present the fluctuations of galaxy number counts within linear perturbation theory, concentrating on the redshift-space 2-point correlation function. In \Cref{sec:flagship_specs} we present the relevant quantities from the \Euclid Flagship simulations used in this work. In \Cref{sec:method} we explain the methods used in our analysis, which are the Fisher matrix and Markov chain Monte Carlo (MCMC). In \Cref{sec:results} we discuss our results and present the method to de-bias the analysis, and in \Cref{sec:con} we conclude. Some details and complementary results are delegated to several appendices.

\textbf{Notation:} In this paper, scalar metric perturbations are described via the gauge-invariant dimensionless Bardeen potentials, $\Phi$ and $\Psi$. In longitudinal gauge the perturbed metric is
\be
\de s^2 = a^2(\eta)\left[-(1+2\Psi)\,c^2\de\eta^2+(1-2\Phi)
\,\delta_{ij} \,\de x^i\de x^j\right] \,,
\ee
where we use the Einstein summation convention over repeated indices.
Here $a(\eta)$ is the cosmic scale factor evaluated at conformal time $\eta$, and $c$ is the speed of light.
In the above, as well as subsequent equations, a prime denotes the derivative with respect to conformal time, {and $\HH = a'/a =Ha$ denotes the conformal Hubble parameter. We normalize the scale factor to $1$ today, $a_0=1$ such that $\HH_0=H_0$}.

\section{Fluctuations of spectroscopic galaxy number counts}

\subsection{Galaxy number counts}
An important observable of the \Euclid satellite will be the galaxy number counts, i.e.,\ the number of galaxies $\diff{}N(\bn,z)$ detected in a given small redshift bin $\diff{}z$ around a redshift $z$ and a small solid angle $\diff{}\angle$ around a direction $\bn$.
Expressing $\diff{}N(\bn,z)=n(\bn,z)\,\diff{}z\,\diff{}\angle$ in terms of the angular-redshift galaxy density $n(\bn,z)$ and subtracting the mean
\be
\bar n(z) = \frac{1}{4\pi}\int_\angle n(\bn,z)\; \diff{}\angle \,,
\ee
we define the galaxy number count fluctuation as
\be
\De(\bn,z)  =\frac{n(\bn,z)-\bar n(z)}{\bar n(z)} \,.
\ee
This quantity and its power spectra have been calculated at first order in cosmological perturbation theory in~\citet{Yoo:2009au,Yoo:2010ni,Bonvin:2011bg,Challinor:2011bk,Jeong:2011as}. As it is an observable, the result is gauge invariant.
It is not simply given by the density fluctuation on the constant redshift hypersurface, but also contains volume distortions. Most notable of those are the radial volume distortion from peculiar velocities, the so-called redshift space distortions~\citep[RSD,][]{Kaiser1987}, but also the transversal volume distortion due to weak lensing magnification~\citep{Matsubara:2004fr} and the large-scale relativistic effects identified for the first time in the above references.
The final formula, including the effect of evolution bias~\citep{Challinor:2011bk,Jeong:2011as}, is given by~\citep{DiDio:2013bqa}
{
\begin{align}
\De(\bn,z,m_*) &= b\,{\delta} +\frac{1}{\HH}
\dd_r\left(\bV\cdot\bn\right) {\big|_{\rsource}} \nonumber
\\
&
-\frac{2-5s}{2\rsource}\int_0^{\rsource}\hspace{-0.3mm}\diff r \frac{\rsource-r}{r}\,
\laplacian_\angle (\Phi+\Psi)  +\  \ldots \label{e:DezNF}
\end{align}
Here {$r = r(z)$ is the comoving distance evaluated at redshift $z$}, $\rsource$ is the comoving distance between the observer and the source, $b =\,b(z,m_*)$ is the linear bias of galaxies with magnitude below $m_*$, the magnitude limit of the survey, {$\delta$} is the gauge-invariant density fluctuation representing the density in comoving gauge,
{$\partial_r$ is the derivative w.r.t. the comoving distance $r$, $\bV$ is the velocity of sources in the longitudinal gauge},
and $\laplacian_\angle$ denotes the angular part of the Laplacian.}

The function $s =\,s(z, m_*)$ is the local count slope needed to determine the magnification bias.
The local count slope depends on the magnitude limit $m_*$ of the survey and is given by~\citep[see, e.g.,][]{Challinor:2011bk}
\be
s(z,m_*) = \frac{\dd\log_{10} \bar N(z,m<m_*)}{\dd m_*}\,.
\label{eq:mag_bias_def}
\ee
Here $\bar{N}$ is the cumulative number of objects {brighter than} the magnitude cut $m_*$ (for a magnitude-limited sample).
The first line of \Cref{e:DezNF} corresponds to the standard terms of density fluctuations and redshift space distortions while the second line is the lensing magnification which is the subject of the present paper, see~\citet{Bonvin:2011bg,Challinor:2011bk,Jeong:2011as,DiDio:2013bqa} for details.
\footnote{Note that to obtain \Cref{e:DezNF} we assume that galaxies obey the Euler equation, i.e.,\ that dark matter does not interact and exchange energy or momentum with other constituents~\citep{Bonvin:2018ckp}, but we have not used Einstein's equations.}
The dots at the end of \Cref{e:DezNF} stand for the ``large-scale relativistic terms'' which we omit in our analysis.
These terms are suppressed by factors $\la \HH\,c^{-1}$, where $\la$ is the comoving wavelength of the perturbations, see, e.g.,~\citet{DiDio:2013bqa,Jelic-Cizmek:2020pkh,Euclid:2021rez}.
It is well known that the large-scale relativistic terms are relevant only at very large scales and do not significantly impact the even multipoles of the correlation function on sub-Hubble scales~\citep{Lorenz:2017iez, Yoo:2009au, Bonvin:2011bg}. Some of these relativistic terms will, however, be detectable by measuring odd multipoles in the correlation of two different tracers~\citep[see][]{Bonvin:2013ogt,Gaztanaga:2015jrs,Lepori:2019cqp,Beutler:2020evf,Saga:2021jrh,Bonvin:2023jjq}.
{A detailed study of the signal-to-noise ratio of all relativistic effects in simulated mock catalogues adapted to \Euclid's spectroscopic survey will be presented in~Euclid collaboration: Elkhashab et al. (in preparation).}

The goal of this paper is to study the impact of lensing magnification on the two-point correlation function. Note that the impact of lensing magnification on the angular power spectrum, $C_\ell(z_1,z_2)$, has already been computed and found to be relevant for \Euclid's photometric sample~\citep{Euclid:2021rez}.
However, the angular power spectrum, $C_\ell(z_1,z_2)$ is not well suited to a survey with spectroscopic resolution of $\delta z\lesssim 10^{-3}$ since we would have to split the redshift interval into more than 1000 bins in order to fully profit from the redshift resolution of a spectroscopic survey. This would not only significantly increase the computational effort, but also lead to large shot noise in the auto-correlation spectra.\footnote{While there are methods that address this issue~\citep[see for instance][]{Camera_2018}, we do not make use of them in this paper.} These are the main reasons that, for spectroscopic surveys, the correlation function is a more promising summary statistics than the angular power spectrum and we therefore need to determine the impact of lensing magnification on this statistic.

\subsection{The 2-point correlation function}
\label{2pf}

In spectroscopic surveys, there are two standard estimators used to extract information from galaxy number counts: the 2-point correlation function (2PCF), and its Fourier transform, the power spectrum. In this paper, we concentrate on the correlation function, since lensing magnification can be included in this estimator in a straightforward way. This is not the case for the power spectrum, which requires non-trivial extensions to account for magnification~\citep[see, e.g.,][]{Grimm_2020,Castorina:2021xzs}.

The 2-point correlation function can be calculated in the curved-sky, i.e., without assuming that the two directions $\bn$ and $\bn'$ are parallel
\begin{align}
\xi(\bn, z, \bn', z')=\langle \Delta(\bn, z)\Delta(\bn', z') \rangle \, .
\end{align}
The curved-sky density and RSD contributions were first derived in~\citet{Szalay:1997cc} and~\citet{Szapudi:2004gh}. This method, which can be straightforwardly applied to any local contribution of $\Delta(\bn, z)$, cannot be used to calculate contributions from integrated effects like lensing magnification, which is the main subject of this work. The magnification contribution was calculated in~\citet{Tansella_2018} using an alternative method proposed in~\citet{Campagne_2017}. The detailed expressions for all contributions can be found in~\citet{Tansella_2018}; for completeness, we repeat them in \Cref{sec:fullsky_expressions}.
The expressions for the 2PCF significantly simplify in the flat-sky approximation: in this approximation, one assumes that the two directions $\bn$ and $\bn'$ are parallel, and one neglects the redshift evolution of $\Delta$. In this case, the density and RSD contributions (hereafter called standard terms) in a `thick' redshift bin with mean redshift $\bar{z}$ take the following simple form
\begin{align}
 \xi^\std(d,\bar{z},\mu)=\xi^\std_0(d,\bar{z})+\xi^\std_2(d,\bar{z})L_2(\mu)+\xi^\std_4(d,\bar{z})L_4(\mu)\, ,
\end{align}
where $d$ denotes the comoving separation between the correlated volume elements or `voxels', $\bar{z}$ is the {centre of the bin interval} in which the correlation function is measured, $\mu$ is the cosine of the angle between the direction of observation $\bn$ and the vector connecting the voxels, and $L_\ell$ denotes the Legendre polynomial of order $\ell$.
The standard multipoles, $\xi_\ell^\std$, are given by the well-known expressions
\begin{align}
\begin{aligned}
\xi^{\std}_0(d,\bar{z})&=\left[b^2(\bar{z}) +\frac{2}{3}b(\bar{z})f(\bar{z})+\frac{1}{5}f^2(\bar{z})\right]\mu_0(d,\bar{z})\,,\\  \xi^{\std}_2(d,\bar{z})&=-\left[\frac 43 f(\bar{z})b(\bar{z})+\frac 47 f^2(\bar{z})\right]\mu_2(d,\bar{z})\,,\\
\xi^{\std}_4(d,\bar{z})&=\frac{8}{35}f^2(\bar{z}) \; \mu_4(d,\bar{z})\,,\label{eq:mult}
\end{aligned}
\end{align}
where
\begin{align}
\ng{f(\bar{z}) :=\frac{\mathrm d\ln\delta}{\mathrm d\ln a}}\,,
\end{align}
is the growth rate of structure.
The functions $\mu_\ell(d,\bar{z})$ are given by
\begin{align}
\mu_\ell(d,\bar{z})=\frac{1}{2\pi^2}\int_0^\infty \diff k \; k^2 P_{\delta\delta}(k,\bar{z})j_\ell(kd) \, , \label{eq:mufunction}
\end{align}
where $j_\ell$ denotes the spherical Bessel function of order $\ell$, and $P_{\delta\delta}(k, \bar{z})$ is the linear matter power spectrum in the comoving gauge.
We see that, in the flat-sky approximation, density and RSD are fully encoded in the first three even multipoles.
{The expressions for computing higher-order multipoles without using the flat-sky approximation can be found in \Cref{sec:fullsky_expressions}.}

The magnification contribution can also be simplified using the flat-sky approximation and the Limber approximation. It has been derived in detail in~\citet{Tansella_2018}. It contains an infinite series of multipoles,
\begin{align}
\xi^\magn(d,\bar{z},\mu)=\sum_\ell \xi^\magn_\ell(d,\bar{z})L_\ell(\mu)\, ,
\end{align}
with
\begin{align}
\label{eq:multipoles_lensing}
\xi_\ell^\magn(d,\bar{z})&= \frac{2\ell+1}{2}
\Bigg\{\frac{3\Omm}{2\pi}(1+\bar z)b(\bar z)(5s-2)
d\int_0^1 \diff\mu \; \mu\; L_\ell(\mu) \nonumber
\\
&\times\int_0^\infty \diff k_\perp \; k_\perp \frac{H_0^2}{c^2}
P_{\delta\delta}(k_\perp, \bar{z})J_0\left(k_\perp d\sqrt{1-\mu^2}\right)\\
&+\frac{9\Omm^2}{4\pi}(5s-2)^2\int_0^{\bar{r}} \diff r'\;\frac{(\bar{r}-r')^2r'^2}{\bar{r}^2 a^2(r')}\int_0^1 \diff\mu \; L_\ell(\mu)\nonumber\\
&\times \int_0^\infty \diff k_\perp \; k_\perp \frac{H_0^4}{c^4}
P_{\delta\delta}\left(k_\perp, z(r')\right)J_0\left(k_\perp\frac{r'}{\bar{r}} d\sqrt{1-\mu^2}\right)\Bigg\}\, .\nonumber
\end{align}
Here $H_0 = 100\,h\,\kmsMpc$,
{$\boldsymbol{k}_\perp = \boldsymbol{k} -k\mu \bn$ is the projection of the Fourier space wavevector $\boldsymbol{k}$ to the direction normal to $\bn$}, and $J_0$ denotes the  Bessel function of {zeroth} order. The first two lines of \Cref{eq:multipoles_lensing} contain the density-magnification correlation: when computing the multipoles, one averages over all orientations of the pair of voxels. {F}or each orientation, the galaxy which is further away is lensed by the one in the foreground. This effect clearly has a non-trivial dependence on the orientation angle $\mu$ which enters in the argument of the Bessel function $J_0$. The last two lines contain the magnification-magnification correlation, due to the fact that both galaxies are lensed by the same foreground inhomogeneities. In all the terms, the functions are evaluated at the mean redshift of the bin $\bar{z}$, $\bar{r}$ denotes the comoving distance at that redshift, and $a(r)$ denotes the scale factor evaluated at comoving distance $r$.
For completeness, in \Cref{sec:flatsky_expressions} we list the semi-analytic expressions which allow us to efficiently evaluate the flat-sky magnification terms.

\begin{figure}
  \includegraphics[width=\linewidth]{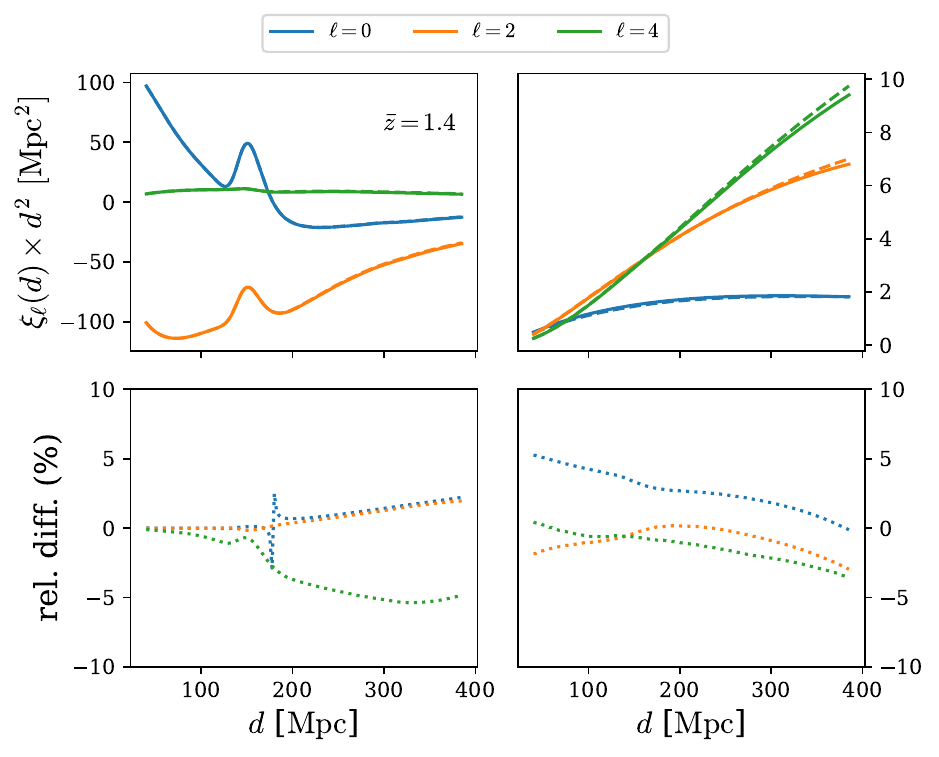}
    \caption{\textit{Top}: comparison of curved-sky (solid) vs. flat-sky (dashed) multipoles, with contributions from standard terms (left) and from just lensing magnification (right). \textit{Bottom}: their relative difference in percent {taking the curved-sky case as the reference value}.}
  \label{fig:comparison_fullsky_flatsky}
\end{figure}

In \Cref{fig:comparison_fullsky_flatsky} we show a comparison between the curved-sky expression and the flat-sky approximation for the standard multipoles (left panel) and the magnification multipoles (right panel), in {one of the} redshift bin{s} of \Euclid, $\bar z=1.4$.
We have checked that most of the constraining power comes from {standard terms of} the monopole and quadrupole below $d=150$\,Mpc, where the difference between the curved-sky and the flat-sky expressions is less than 0.2\% (it reaches 0.7\% for the hexadecapole).
Similar results are obtained for the other redshift bins, hence using the flat-sky approximation is very well justified.
For the magnification contribution to the monopole, we see that the flat-sky approximation differs from the curved-sky result already at small separation by roughly 5\%. However, since the magnification is a sub-dominant contamination to the total signal, a 5\% error is perfectly acceptable.
{Note that the difference between curved-sky and flat-sky actually increases for small separations; as shown in~\citet{Jelic_Cizmek_2021}, this can occur when the dominant contribution to the multipoles is the density-lensing term, for which the accuracy of the flat-sky approximation becomes progressively worse at smaller scales.}

\begin{figure}
  \includegraphics[width=\linewidth]{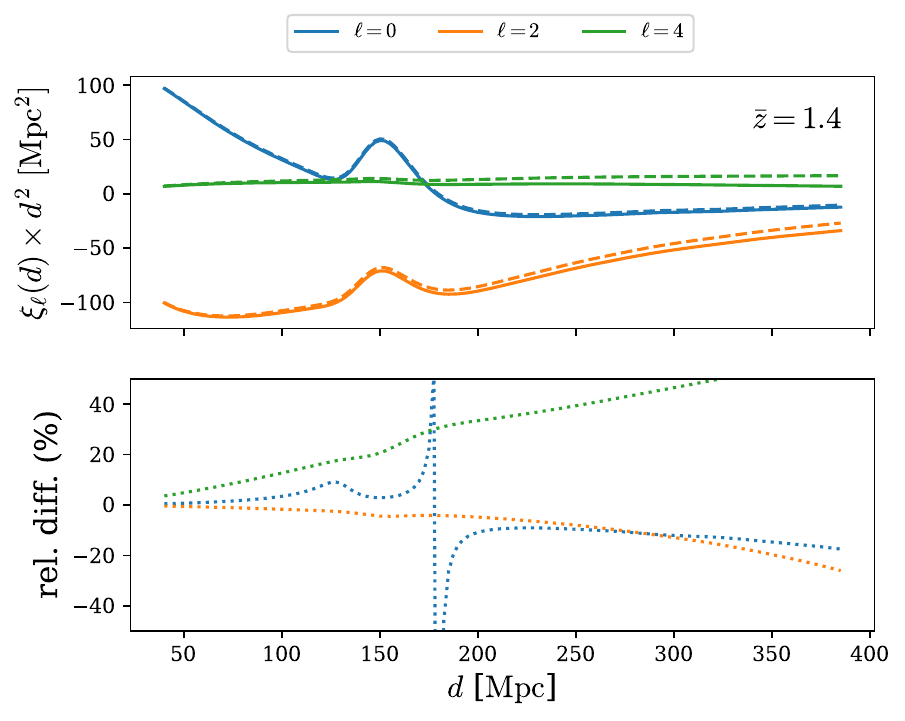}
    \caption{\textit{Top}: comparison of multipoles without (solid) and with (dashed) lensing magnification. \textit{Bottom}: their relative difference in percent {taking the case with magnification as the reference value}.}
  \label{fig:comparison_lensing_vs_no_lensing}
\end{figure}

In \Cref{fig:comparison_lensing_vs_no_lensing} we compare the 2PCF with and without lensing magnification. On small scales, lensing magnification is not very important. However, for scales of more than 300 Mpc it can contribute up to 20\% to the monopole and the quadrupole and up to 50\% or more to the hexadecapole. We see that the further apart the galaxies are, the more significant is the contribution from lensing magnification to their correlation. This is due to the fact that the density correlations quickly decrease with separation, while the lensing magnification correlations do not. The magnification-magnification correlations are indeed integrated all the way from the sources to the observer, and therefore contain contributions from small scales, when the two lines of sight are close to the observer.

We now  assess the impact of these multipoles on the analysis of data from \Euclid. In particular, we  determine the following: first, if the contribution from magnification can improve our measurement of cosmological parameters; and second, {to which extent neglecting magnification in the analysis will} shift the best-fit value of the parameters, consequently biasing the analysis.

We consider two cases.
In the first case, we fix the cosmological model, and we study how magnification impacts the parameters of this model. For this case, we will study two models: a minimal $\Lambda$CDM model and a dynamical dark energy model. In the second case, we perform a model-independent analysis, i.e.,\ we rewrite the standard multipoles in terms of the power spectrum at $z_*$, {using that $P_{\delta\delta}(\bar{z})=P_{\delta\delta}(z_*)\sigma^2_8(\bar{z})/\sigma^2_8(z_*)$. We} choose $z_*$ to be well in the matter-dominated era, before acceleration started.
We assume that, at $z_*$, general relativity is valid, and that the power spectrum is therefore fully determined by the early Universe parameters that have been measured by the CMB. {With this, the functions $\mu_{\ell}$ depend only on the power spectrum at $z_*$, while the evolution from $z_*$ to $\bar{z}$ is fully encoded in two functions:
\begin{align}
\tilde{f}(\bar{z})=f(\bar{z})\,\sigma_8(\bar{z})\quad\mbox{and}\quad \tilde{b}(\bar{z})=b(\bar{z})\,\sigma_8(\bar{z})\, .
    \label{eq:tilde_definition}
\end{align}
We obtain for} the multipoles of the standard terms
\begin{align}
\begin{aligned}
\xi^{\std}_0(d,\bar{z})&=\left[\tilde{b}^2(\bar{z}) +\frac{2}{3}\tilde{b}(\bar{z})\tilde{f}(\bar{z})+\frac{1}{5}\tilde{f}^2(\bar{z})\right]
\frac{\mu_0(d,z_*)}{\sigmastarsquared}\,,\\
\xi^{\std}_2(d,\bar{z})&=-\left[\frac 43 \tilde{f}(\bar{z})\tilde{b}(\bar{z})+\frac 47 \tilde{f}^2(\bar{z})\right]
\frac{\mu_2(d,z_*)}{\sigmastarsquared}\,,\\
\xi^{\std}_4(d,\bar{z})&=\frac{8}{35}\tilde{f}^2(\bar{z})\,
\frac{\mu_4(d,z_*)}{\sigmastarsquared}\,.
\end{aligned}
\label{eq:mult_z*}
\end{align}

In this case the functions $\mu_\ell(d,z_*)$ are considered fixed, since they are very well determined by CMB measurements, and the functions $\tilde{f}$ and $\tilde{b}$ are two free functions that depend on the mean redshift $\bar{z}$ of the bins, see also~\citet{Jelic-Cizmek:2020pkh} for an introduction of this method. These two free functions fully encode any deviations from general relativity at late time. The only approximation that enters here is that we neglect the $k$-dependence of the growth of density such that $\sigma_8$ and $f$ depend only on redshift. This assumption can easily be relaxed: it slightly complicates the analysis, since $\tilde{f}$ and $\tilde{b}$ would have to be taken inside the integrals in \Cref{eq:mufunction}, but it does not change the procedure.

This model-independent analysis is one of the key goals of the \Euclid spectroscopic survey. It is very powerful, since it allows us to measure the growth rate of structure without assuming a particular model of gravity or dark energy. This growth rate can then be compared with the predictions of any model beyond $\Lambda$CDM. In the following, we will determine how neglecting magnification in the analysis could shift the best-fit values of $\tilde{b}$ and $\tilde{f}$ in each redshift bin. Note that here for simplicity we fix the cosmological parameters that determine the functions $\mu_\ell$ at early time, $z_*$, to their fiducial value extracted from \textit{Planck} data~\citepalias{2016A&A...594A..14P}. In practice, one can also let these parameters vary and perform a combined analysis with CMB data.

It is worth mentioning that, by construction, the analysis using multipoles of the correlation function does not account for correlations between different redshift bins: the correlation function is averaged over directions within a given bin, and each redshift bin is considered to be independent.
However, the main motivation of measuring the multipoles of the correlation function is to extract the growth rate $f$, which is encoded in the peculiar velocities of galaxies within linear perturbation theory.
Moreover, since the correlations of peculiar velocities quickly decrease with separation, one does not lose a significant amount of information by neglecting cross-correlation between bins.
The situation is of course different for magnification, which, as it is an integral along the line-of-sight, is strongly correlated between different bins. Therefore, we expect that {neglecting cross-correlations of different bins}  will strongly reduce the magnification signal, compared to the angular power spectra used in the analysis of the photometric sample, which is able to account for correlations between the bins~\citep{Euclid:2021rez}. As such, the shift induced by neglecting lensing magnification is expected to be smaller in the spectroscopic analysis than in the photometric one.

\section{\Euclid specifics from the Flagship simulation}
\label{sec:flagship_specs}
\begin{figure}
  \includegraphics[width=\linewidth]{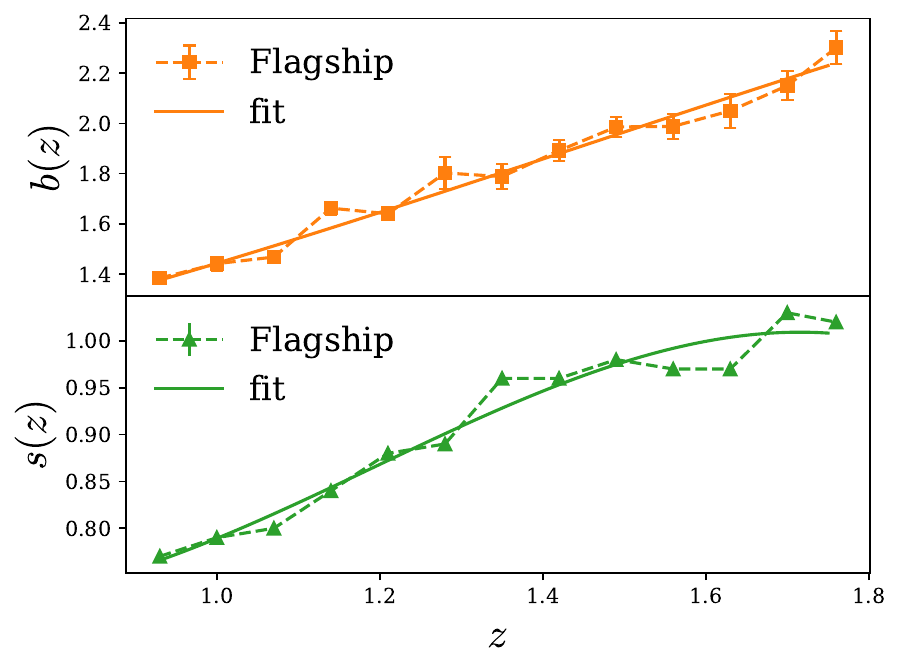}
    \caption{We show the galaxy bias (top panel) and the local count slope (lower panel), with linear interpolation (dashed), as used in our analysis, along with their associated error bars, {as well as a polynomial fit (solid); for the exact numerical values of the coefficients, refer to \Cref{eq:fit_biases}.}}
\label{f:b_and_s}
\end{figure}

In order to calculate the linear galaxy bias and local count slope observables for this analysis, which we will use as our fiducial values for the Fisher and MCMC analyses, we use Flagship \texttt{v1.8.4} galaxy mock samples, whose redshift distribution of the number density, $\numberdensity(z)$, is split in 13 {equally spaced} bins {(in redshift)}, between $z=0.9$ and $z=1.8$, in real space.
As we see from \Cref{f:b_and_s}, this allows us to accurately capture the redshift evolution of the bias and the local count slope, while still having enough galaxies in each bin to obtain a precise measurement.
We impose a cut in the H${\alpha}$ flux $F_{\text{H}{\alpha}}$ (in units of $\text{erg}\,\text{s}^{-1}\,\text{cm}^{-2}$), $\log_{10}{[F_{\text{H}{\alpha}}/(1\, \text{erg}\,\text{s}^{-1}\,\text{cm}^{-2})]} > -15.7$, that can be transformed to the corresponding AB magnitude limit, $m_* = m_\text{AB}<-15.75$.\footnote{%
 We note that the $\numberdensity(z)$ include survey specific effects, such as purity and completeness following the pipeline of the Flagship Image Simulations.
We have observed that not considering these two systematics can affect the linear galaxy bias value up to a $5\%$ depending on the redshift, while the magnification bias is not significantly affected by this.}

\subsection{Linear galaxy bias}
The linear galaxy bias is obtained by fitting the curved-sky angular power spectrum of the data to the corresponding fiducial prediction for {$\numberdensity(z)$ in each redshift bin}.
The angular power spectrum is obtained using \texttt{Polspice}\footnote{\url{http://www2.iap.fr/users/hivon/software/PolSpice/}.} with a mask to generate 100 jackknife regions that we use to calculate the covariance matrix.
{This masked region corresponds to the area of the sky that the Cosmohub's Flagship \texttt{v1.8.4} release~\citep{TALLADA2020100391,2017ehep.confE.488C} does not cover, which corresponds to {7/8} of the total sky. The mask is also generated with \texttt{Polspice} by masking the pixels outside the region $0\degree<\text{RA}<90\degree$ and $0\degree<\text{Dec}<90\degree$. Then when considering the jackknife regions, we use a k-means clustering algorithm to select 100 regions with roughly the same number of pixels (since each \texttt{HEALPix}\footnote{\url{https://healpix.sourceforge.io/}.} pixel covers the same sky area) inside the unmasked region. For each jackknife resampling we include into the mask the pixels of a different region in order to exclude it from that jackknife iteration $C_\ell$'s calculation.}
The prediction is determined using \texttt{CCL}\footnote{\url{https://github.com/LSSTDESC/CCL}.}~\citep{ccl_paper} with the fiducial cosmological parameter values being those used in the Flagship \texttt{v1.8.4} simulation, see \Cref{tab:fid} below.
The linear scales considered range from $\ell=50$ to an $\ell_\text{max}$ that increases as we go to higher redshifts (as the effective non-linear galaxy bias scale shifts to higher multipoles, i.e, smaller angular scales), starting at $\ell_\text{max}\sim300$ for $z=0.9$ to $\ell_\text{max}\sim500$ for $z=1.8$.
Since \texttt{CCL} does not yet allow us to perform calculations without the Limber approximation, we employ it for all of the scales used to estimate the linear galaxy bias. {We set the minimum multipole to $\ell=50$ which is a rather conservative limit in order to avoid large Limber approximation deviations from the theory at any redshift considered for this analysis. On top of that, scales with $\ell<50$ usually have very high errorbars due to sample variance so they can be ignored since their statistical weight to estimate the galaxy bias is very low. The maximum scale for each redshift is estimated by comparing the relative ratio between the linear and non-linear matter power spectrum prediction and setting a maximum relative difference of 3$\%$. This maximum difference should be good enough since galaxy clustering is known to follow linear predictions down to smaller scales than dark matter.}
 At the scales used, the Limber approximation adds up to a 4\% variation of the predicted angular power spectrum which translates into an error below 2\% on the galaxy bias estimation.
The $\chi^2$ distribution is then calculated for different values of the galaxy bias in relation to the square ratio of the data $C_\ell$'s to the prediction $C_\ell$'s using only the diagonal values of the jackknife covariance matrix.
We estimate the linear galaxy bias as the minimum value of the $\chi^2$ distribution, and we set the error to the $1\sigma$ variance.

\subsection{Local count slope}

To measure the local count slope from the Flagship catalogues, we compute, in each of the 13 redshift bins, the cumulative number of galaxies $\bar{N}(z, m<m_*)$ at the magnitude limits $m_*$ and $m_*\pm 0.04$, and we compute the logarithmic derivative to obtain $s(z,m_*)$ through \Cref{eq:mag_bias_def}. We also generate 100 jackknife regions in order to calculate the variance of the results.
In \Cref{f:b_and_s}, we see that $s$ increases with redshift; this is due to the fact that a fixed apparent magnitude threshold, $m_*$, corresponds to a larger intrinsic luminosity threshold $L_*$ at high redshift than at low redshift.
This is due to the fact that the slope in $m$ of the Schechter luminosity function, which is assumed here, increases with redshift.

\section{Method}
\label{sec:method}
This study employs two complementary approaches for forecasting the constraining capabilities of future \Euclid data: the Fisher matrix formalism and the Markov chain Monte Carlo method. The details of each approach are outlined in \Cref{sec:fisher}, \Cref{sec:fisher-bias}, and \Cref{sec:mcmc-intro}, respectively.
{We use the code \coffe{}\footnote{Available at \url{\coffeurl}.}, which has been validated against the code \cbl{}\footnote{Available at \url{\cblurl}. In this work, we use \texttt{git} revision \texttt{7f08f470e0} of the code.} (for details of the validation, see \Cref{sec:validation}), to compute the multipoles of the 2PCF.}

\subsection{The Fisher matrix formalism: Cosmological constraints}
\label{sec:fisher}
The Fisher matrix can be defined as the expectation value of the second derivatives of the logarithm of the likelihood under study with respect to the parameters of the model~\citep[see, e.g.,][]{Tegmark1997}:
\begin{equation}
    \fisher_{\alpha\beta}=\Braket{-\frac{\partial^2\,\text{ln}\mathcal{L}}{\partial \theta_{\alpha} \partial \theta_{\beta}}}\,,
\end{equation}
where $\alpha$ and $\beta$ label two model parameters $\theta_{\alpha}$ and $\theta_{\beta}$.

In the particular case of Gaussian-distributed data, the Fisher matrix is given by
\begin{equation}\label{eq:fisher}
    \fisher_{\alpha\beta}=\frac{1}{2}\,\text{Tr}\left[\frac{\partial \tens{C}}{\partial \theta_{\alpha}}\tens{C}^{-1}\frac{\partial \tens{C}}{\partial \theta_{\beta}}\tens{C}^{-1}\right]+\sum_{pq}\frac{\partial {\tens{D}_p}}{\partial \theta_{\alpha}}\left(\tens{C}^{-1}\right)_{pq}\frac{\partial {\tens{D}_q}}{\partial \theta_{\beta}}\,,
\end{equation}
where {$\tens{D}$} represents the mean of the data vector and $\tens{C}$ is the covariance matrix of the data. The trace, Tr, and the sum over the indexes $p$ and $q$ stand for summation over the different elements of the data vector.

In the present analysis we consider the 2-point correlation function as our main observable. Given the set of model parameters $\{\theta_\alpha\}$, the Fisher matrix for the multipoles of the 2-point correlation function measured in a bin centered in $\bar{z}_i$ is
\be
\fisher^\text{bin}_{\alpha\beta}(\bar{z}_i) =  \sum_{jk} \sum_{\ell m} \frac{\partial \xi_\ell(d_j, \bar{z}_i)}{\partial \theta_\alpha} \,\,\covariance^{-1}\left[\xi^j_{\ell},\xi^k_{m}\right] (\bar{z}_i)\,\, \frac{\partial \xi_{m}(d_k, \bar{z}_i)}{\partial \theta_\beta}\,, \label{eq:fisher-bin}
\ee
where the sum runs over the voxel separations $\{d_j, d_k\}$ as
well as the even multipoles $\ell, m = 0, 2, 4$ {and $\xi^j_{\ell}\equiv\xi_{\ell}(d_j)$}.
Note that (angular) power spectra observables follow a Wishart distribution if fluctuations are Gaussian. In this case, the Fisher analysis gives a better approximation if we consider only the second term in \Cref{eq:fisher}\,\citep[see, e.g.,][]{2013A&A...551A..88C, Bellomo:2020pnw}. In the following we assume that the same is true for the multipoles of the correlation function.
The binned covariance of the 2PCF multipoles at mean redshift $\bar{z}_i$, denoted with $\covariance$, is computed following the Gaussian theoretical model described in~\citet{Grieb:2015bia} and~\citet{Hall:2016bmm}. The cosmic variance contribution includes only the density and RSDs, while magnification is neglected.
{This is a good approximation, since the covariance is a four-point function, which contains a sum over all possible separations between pairs of pixels. Hence even at large separation, the covariance is dominated by correlations at small scales, where density and RSD strongly dominate over magnification.}
The shot noise contribution is estimated from the number densities reported in~\citet{Blanchard:2019oqi}~\citepalias[henceforth referred to as][]{Blanchard:2019oqi}, Table 3.
The full expression for the covariance can be found in \Cref{ap:cov}.

Following~\citetalias{Blanchard:2019oqi}, we neglect the cross-correlations between redshift bins. Thus, the full Fisher matrix is
\be
\fisher_{\alpha\beta} = \sum_{\bar{z}_i}  \fisher^\text{bin}_{\alpha\beta} (\bar{z}_i) \,.\label{eq:fisher-full}
\ee
This approximation is justified by the high precision of spectroscopic redshift estimates, which leads to essentially no overlap between redshift bins. In the case of a photometric analysis, cross-correlations between different bins can provide significant information\,\citep[see, e.g.,][]{Tutusaus_2020,Euclid:2021rez}.
The marginalized 1$\sigma$ errors on the cosmological parameters can then be estimated from the Cramér-Rao bound, that is
\be
\sigma_{\alpha}= \sqrt{(\fisher^{-1})_{\alpha\alpha}}.
\ee

It is important to mention that, although the Fisher matrix formalism is a powerful forecasting tool, some limitations do exist. A Fisher forecast uses a Gaussian approximation by construction, which can differ from the true posterior if the data are not constraining enough. Furthermore, the signal and covariance may have a strong non-linear dependence on the parameters $\{\theta_\alpha\}$, in which case the Fisher matrix does not capture all of the information about the likelihood. In order to validate the results of our Fisher formalism, we also perform, for one of the cases, a Markov chain Monte Carlo analyses to properly sample the posterior of the parameters (see \Cref{sec:mcmc-intro}). As we will see, we find that the relevance of lensing magnification is well captured by a Fisher forecast.
Another drawback of the Fisher formalism worth mentioning is that it only provides forecast uncertainties around a fiducial model. In this analysis we are also interested in the bias on the posteriors because of wrong model assumptions (neglecting magnification). The standard Fisher formalism prevents us from doing this study, but extensions to the formalism can be considered, as described in \Cref{sec:fisher-bias}.
{All Fisher forecasts were computed with the Python package~\fitk.\footnote{Available at \url{\fitkurl}.}}

\subsection{The Fisher matrix formalism: bias on parameter estimation}\label{sec:fisher-bias}
The Fisher matrix formalism described above allows us to quantify the gain or loss in constraining power, when magnification is included in the theoretical model for the observed multipoles of the 2PCF.
This study can be carried out by simply comparing the Fisher matrix in \Cref{eq:fisher-full} and the corresponding marginalized constraints when magnification is neglected or included in the analysis.

Another, actually more important question to address is whether neglecting magnification leads to significant
biases (shifts) in the inferred cosmological parameters. In order to answer this question, we follow the approach described for example in~\citet{Taylor:2006aw} and widely adopted in the literature, see~\citet{Kitching:2008eq, Camera:2014sba, DiDio:2016ykq, Cardona:2016qxn, Lepori:2019cqp, Jelic-Cizmek:2020pkh, Euclid:2021rez}.
We extend the parameter space to include the amplitude of magnification $\epsilon_\mathrm{L}$.
We can explicitly write the dependence of our model on $\epsilon_\mathrm{L}$ as follows
\be
\xi_\ell(d_j, \bar{z}_i)  = \xi_\ell^\std (d_j, \bar{z}_i)  + \epsilon_\mathrm{L} \xi_\ell^\magn (d_j, \bar{z}_i)\,,
\ee
where $\xi_\ell^\std$ represents the standard contributions of density and RSD, while $\xi_\ell^\magn$
is the magnification contribution, which includes the terms magnification $\times$ magnification and the cross-correlation of magnification with density and RSD.
The amplitude $\epsilon_\mathrm{L}$ is not a free parameter, but rather a fixed one, that is set to either $0$ (in a `wrong' model that neglects magnification), or $1$ (in a `correct' model that consistently includes magnification).
The `wrong' and `correct' models share a common set of parameters, $\{\theta_\alpha\}$, and their estimation will be biased in the wrong model as a result of the shift in the fixed parameter $\epsilon_\mathrm{L}$.
Using a Taylor expansion of the likelihood around the wrong model, and truncating the series at the linear order, we obtain
the following formula for the biases
\begin{equation}
\Delta (\theta_\alpha) = \sum_{\beta} \left(\fisher^{-1}\right)_{\alpha\beta} B_\beta\,, \label{eq:shift}
\end{equation}
where $\fisher$ is the Fisher matrix of the common set of parameters evaluated for the wrong model, and
\begin{equation}
B_\beta = \sum_{i} \sum_{jk} \sum_{\ell m} \xi_\ell^\magn (d_j, \bar{z}_i) \,\,\covariance^{-1}\left[\xi^j_{\ell},\xi^k_{m}\right] (\bar{z}_i)\,\, \frac{\partial \xi_\ell^\std(d_k, \bar{z}_i)}{\partial \theta_\beta}\,. \label{eq:shifts}
\end{equation}
\Cref{eq:shifts} implicitly assumes that magnification constitutes a small contribution to the observable; therefore, the outcome can be quantitatively trusted only when small values of the biases are found.
Nevertheless, large biases are a clear indication that the lensing magnification significantly contributes to the observable and that the starting hypothesis should be rejected.
Therefore, it is a good diagnostic to assess whether magnification can be neglected or should be modeled in the analysis.

\subsection{Markov chain Monte Carlo}
\label{sec:mcmc-intro}
The Markov chain Monte Carlo (MCMC) method is a standard statistical technique where we numerically sample the posterior probability starting from a prior probability and assuming a likelihood function (the probability of the data given the hypothesis). An excellent description of the method can be found in~\citet{Verde}.

In our analysis, under the assumption that our data is Gaussian, we {sample the posterior of} the following likelihood:
\be
\ln{(\mathcal{L})} = - \frac{1}{2} \Delta \boldsymbol{\xi}^T \mathsf{C}^{-1} \Delta \boldsymbol{\xi},
\ee
where $\mathsf{C}$ is the covariance matrix, and $\Delta \boldsymbol{\xi}$ is a vector whose elements are given by
\be
{\Delta}\boldsymbol{\xi}_{[\ell, i, j]} = \xi_\ell(d_j, \bar{z}_i)^\mathrm{ref} - \xi_\ell (d_j, \bar{z}_i),
\ee
where the first term is part of a synthetic data set computed previously used as our `reference' or `fiducial model', while the second term is computed at each steps of the MCMC, varying the value of free parameters inside the parameters space described by the prior function.
To simplify the analysis, we neglect the dependence of the covariance matrix on cosmological parameters, that we fix to their reference values.

We assume a flat prior {density} for each parameter. In order to speed up the convergence of our chains, we assume as free parameters {(with flat priors)} $\omm$ and $\omb$ instead of $\Omm$ and $\Omb$, and subsequently reparametrize the chain using the relation $\omega_{i, 0} = \Omega_{i, 0} h^2$.

We use the Python package \texttt{emcee}\footnote{Available at~\url{https://github.com/dfm/emcee}.}~\citep{emcee} to implement the MCMC. Our sampler is composed of 32 walkers using the ``stretch move'' ensemble method described in~\citet{ensemble_move}. Each walker generates a chain with a number of steps of the order $10^5$ before converging. Our MCMC code is run in parallel using the Python package \texttt{schwimmbad}~\citep{schwimmbad}.

In our analysis we also discard a number of points as burn-in, given by twice the maximal integrated auto-correlation time $\tau$ of all the parameters~\citep{ensemble_move}.\footnote{The time $\tau$ can be considered to be the number of steps that are needed before the chain ``forgets'' where it started.} The results of the sampling are then analysed with the Python package \texttt{GetDist}~\citep{getdist}.

\section{Results}
\label{sec:results}
\begingroup
\newcommand{\zmean}{\bar{z}}

We compute the impact of neglecting magnification for three different cases: a minimal $\Lambda$CDM model, a dynamical dark energy model, and a model-independent analysis measuring the bias and growth rate. For each case, we compute the change in the constraints due to including magnification and the shift in the parameters due to neglecting magnification.

{To be consistent with the Flagship simulation}, the fiducial cosmology adopted in our analysis is a flat $\Lambda$CDM model with no massive neutrino species.
The set of parameters varied in the analysis comprises:
the present matter and baryon density parameters, respectively $\Omm$ and $\Omb$;
the dimensionless Hubble parameter $h$;
the amplitude of the linear density fluctuations within a sphere of radius 8 \hMpc{} at present time, $\sigma_8$;
the spectral index of the primordial matter power spectrum $n_\text{s}$;
and the equation of state for the dark energy component $\{\wzero, \wa\}$, which parametrize the time evolution of the dark energy equation of state parameter as
\begin{equation}
    w(z) = \wzero + \wa \frac{z}{1 + z}.
    \label{eq:dark_energy_eos}
\end{equation}
This model is also known as Chevallier--Polarski--Linder (CPL) parametrization~\citep{Chevallier:2000qy, Linder_2003}.
The fiducial values of the cosmological parameters used in the analysis are reported in \Cref{tab:fid}.
They correspond to the $\wzero\wa$CDM parameters used in the \Euclid Flagship simulation, see \Cref{sec:flagship_specs}.

\begingroup
\setlength{\tabcolsep}{0.5em}
\begin{table}[!ht]
\caption{Fiducial values of the cosmological parameters.}
\begin{center}
    \begin{tabular}{cccccccc}
        \toprule
        $\Omm$ & $\Omb$ & $\sigma_8$ & $\ns$ & $h$ & $\wzero$ & $\wa$ \\
        \midrule
        0.319 & 0.049 & 0.83 & 0.96 & 0.67 & \minus{}1 & 0 \\
        \bottomrule
    \end{tabular}
\end{center}
\label{tab:fid}
\end{table}
\endgroup

In addition to these cosmological parameters, we introduce nuisance parameters and marginalise over them;
in particular, the linear galaxy bias in each redshift bin, $\{b_i\}, \ i = 1, \ldots, N_\text{bins}$, are included as nuisance parameters.
We model them as constant within each redshift bin, and we have estimated their fiducial values using the Flagship simulation, \texttt{v1.8.4}, as described in \Cref{sec:flagship_specs}.
{We list the values of the nuisance parameters used in each redshift bin as well as the expected density of emitters in \Cref{tab:gc_nofz}.}
The impact of magnification on the cosmological parameters may depend on the model chosen to describe our Universe.
We therefore run our analysis for two different cosmological models and comment on the difference between the results when relevant.
We consider:
\begin{enumerate}
\item
A minimal {flat} $\Lambda$CDM model, with five free parameters
$\{\Omm, \Omb, h, \ns, \sigma_8\}$ + nuisance parameters.
\item
A {flat} dynamical dark energy model, with seven free parameters
$\{\Omm, \Omb, \wzero, \wa, h, \ns, \sigma_8\}$ + nuisance parameters.
\end{enumerate}

We also include a cosmology-independent analysis, where as free parameters we consider the modified galaxy bias and the modified growth rate in each redshift bin, $\tilde{f}(z)$ and $\tilde{b}(z)$, defined in \Cref{eq:tilde_definition}.
To be conservative, as well as to reduce the impact of nonlinearities, we only consider separations between $d_\text{min} = 40\text{ Mpc}$ and $d_\text{max} = 385\text{ Mpc}$ in each redshift bin, and, unless specified otherwise, multipoles $\ell \in \{0, 2, 4\}$.
We use voxels of size $L_\text{p} = 5\text{ Mpc}$. We checked that reducing them further to 2.5\,Mpc does not improve the constraints anymore, due to shot noise which saturates the signal-to-noise ratio for too small voxel sizes.

\begin{table*}[ht]
\centering
\caption{%
The expected number density of observed H$\alpha$ emitters for the \Euclid spectroscopic survey, extracted from the Flagship simulation in each redshift bin.
The first two columns show the minimum, $z_\mathrm{min}$, and maximum, $z_\mathrm{max}$, redshift of each bin.
The third column shows the comoving number density of sources, $\numberdensity(z)$; the fourth column lists the total comoving volume of the redshift bin.
The last two columns respectively denote the linear galaxy bias and the local count slope of the sources, evaluated at the mean redshift.
The values of the biases have been obtained by performing a cubic interpolation~\citep[using the \texttt{SciPy} Python package, described in][]{scipy} on the data described in \Cref{sec:flagship_specs}.
}
\label{tab:gc_nofz}
\begin{tabular}{cccccc}
\toprule
$z_{\rm min}$ & $z_{\rm max}$ & $\numberdensity (\bar z)\,[h^3\,\mathrm{Mpc}^{-3}]$ & $V_{\rm s}(\bar z)\,[h^{-3}\,\mathrm{Gpc}^3]$ & $b(\bar z)$ & $s(\bar z)$ \\
\midrule
0.90 & 1.10 & $4.71\times 10^{-4}$ & 7.94   &  1.441  & 0.79 \\
1.10 & 1.30 & $3.75\times 10^{-4}$ & 9.15   &  1.643  & 0.87 \\
1.30 & 1.50 & $2.90\times 10^{-4}$ & 10.05  &  1.862  & 0.96 \\
1.50 & 1.80 & $2.01\times 10^{-4}$ & 16.22  &  2.078  & 0.98 \\
\bottomrule
\end{tabular}
\end{table*}

\endgroup

\subsection{Lensing magnification signal-to-noise}
\begingroup
\newcommand{\zmean}{\bar{z}}
\newcommand\snr{\ensuremath{\text{S} / \text{N}}}
\newcommand{\rmin}{\ensuremath{d_\text{min}}}
As an estimate of the impact of lensing magnification, we first compute the signal-to-noise ratio (\snr{}) of the lensing contribution to the multipoles of the 2PCF, which we define as
\begin{equation}
\snr(\zmean_i)
:=
\sqrt{
\sum\limits_{j, k, \ell, m}
\xi^\magn_{\ell}(d_j, \zmean_i)
\textsf{C}^{-1}\left[\xi^j_{\ell}, \xi^k_{m}\right](\zmean_i)
\xi^\magn_{m}(d_k, \zmean_i)
},
\end{equation}
where the sum goes over all pairs of voxels and all even multipoles taken into consideration.
The results are shown in \Cref{fig:results-snr} and \Cref{tab:snr}. Since lensing magnification also contributes to multipoles larger than the hexadecapole, we show the $\snr$ for two cases: $\ell_\text{m}=4$ and $\ell_\text{m}=6$, {where $\ell_\text{m}$ denotes the highest multipole used in the analysis.}
As we can see, the \snr{} is smallest in the lowest redshift bin, and increases as we go to higher redshifts.
This is a consequence of two effects: first, the local count slope for \Euclid, $s(z)$, increases with increasing redshift (see \Cref{tab:gc_nofz} and \Cref{f:b_and_s}).\footnote{It is important to note that $s(z) > 2/5$ for the redshift range considered in this work, since for $s(z) = 2/5$ the effect of lensing magnification vanishes.} Second, the lensing magnification term is an integrated effect, and as such, has the largest impact at high redshift.

\begingroup
\setlength{\tabcolsep}{0.82em}
\renewcommand{\arraystretch}{1.2}
\begin{table}[htp]
    \caption{The \snr{} per redshift bin of lensing magnification for the configurations with $\ell_\text{m}=4$ and $\ell_\text{m}=6$ respectively.}
    \label{tab:snr}
    \centering
        \begin{tabular}{ccc}
        \toprule
            {$\bar{z}$} & {$\snr(\ell_\mathrm{m}=4)$} & {$\snr(\ell_\mathrm{m}=6)$}\\
            \midrule
            {1.00} & $0.50$ & $0.62$\\
            {1.20} & $0.92$ & $1.11$\\
            {1.40} & $1.52$ & $1.78$\\
            {1.65} & $2.50$ & $2.87$\\
        \bottomrule
        \end{tabular}
\end{table}
\endgroup

\begin{figure}
  \includegraphics[width=\linewidth]{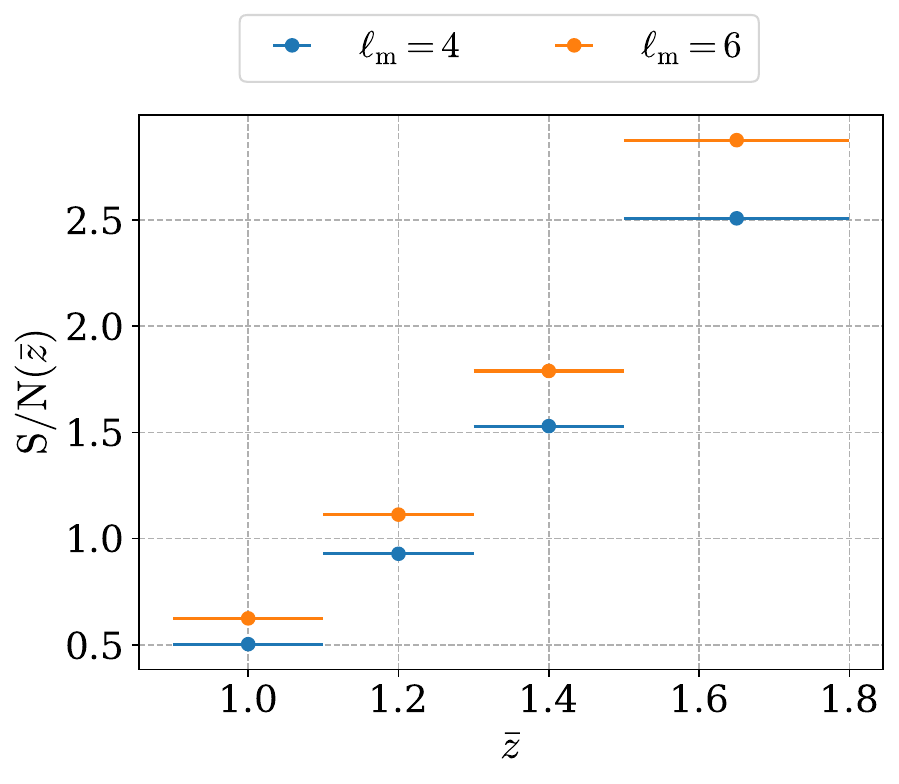}
  \caption{The \snr{} of lensing magnification for 2 different scenarios, with $\rmin = 40\; \text{Mpc}$. The horizontal bars denote the widths of the redshift bins.}
  \label{fig:results-snr}
\end{figure}

\endgroup

\subsection{Full-shape cosmological analysis}
In this section, we study the impact of magnification on the full-shape cosmological analysis of the 2-point correlation function, for the \Euclid{} spectroscopic sample.
Magnification in principle affects both the best-fit estimation of cosmological parameters and their constraints.
In order to quantify the relevance of the effect, we adopt the Fisher formalism described in \Cref{sec:fisher}, and we validate
the results by comparing them to the outcome of a full Markov chain Monte Carlo (MCMC) analysis.
In order to estimate the impact of magnification on constraints on cosmological parameters, we run two Fisher analyses, one with and one without lensing magnification, from which we estimate the marginalized 1$\sigma$ errors, and we compare the values in the two cases.

\begin{table*}[ht]
    \centering
    \caption{
        Impact of magnification on the full-shape analysis.
        The two blocks show our results for the Fisher analysis (top block) and the MCMC analysis (bottom block).
        Fisher analysis: the first two rows show the difference in constraints for a $\Lambda$CDM cosmology (top) and a $\wzero\wa$CDM cosmology (bottom) between the model without lensing magnification, and one with lensing magnification, obtained using the Fisher formalism. First row: in the model with magnification, the local count slope parameters are fixed, that is, the values of $s(z)$ are assumed to be exactly known. Second row: we assume no prior knowledge on the local counts slope, we marginalize over its values in each redshift bin.
        The third row show the shifts in the best-fit estimates due to neglecting magnification in the theory model in $\Lambda$CDM cosmology and a $\wzero\wa$CDM cosmology (bottom).
        MCMC analysis: we show the improvement in cosmological constraints (assuming exact knowledge of the local count slope parameters), and the shifts in the best-fit estimates when magnification is neglected, for a $\Lambda$CDM cosmology.
        \label{tab:fisher_const_lensing_no_lensing}
    }
\adjustbox{max width=\textwidth}{
\begin{tabular}{r r r  r  r  r  r  r  r}
    \toprule
    & \backslashbox{quantity}{$\theta$}    &    $\Omm$ &  $h$  & $\ns$ & $\Omb$ & $\sigma_8$ & $\wzero$ & $\wa$ \\
    \midrule
 \multirow{6}{*}{Fisher~~} &   $1 - \sigma_\magn / \sigma_\text{no \magn}$ ($\Lambda$CDM) (\%)&    4.16 & 4.21 & 0.92 & 5.26 & 5.05 & --- & --- \\
   & $1 - \sigma_\magn / \sigma_\text{no \magn}$ ($\wzero\wa$CDM) (\%)& 0.26 & 1.64 & \minus{}0.26 & 0.34 & 6.30 & 10.20 & 9.90 \\
    \cmidrule{2-9}
  &  $1 - \sigma_{s\text{ marg}}/ \sigma_\text{no \magn}$ ($\Lambda$CDM) (\%)& \minus{}6.29 & \minus{}6.28 & \minus{}1.80 & \minus{}6.22 & \minus{}6.35 & --- & --- \\
  &  $1 - \sigma_{s\text{ marg}} / \sigma_\text{no \magn}$ ($\wzero\wa$CDM) (\%)& \minus{}0.02 & \minus{}0.01 & \minus{}1.87 & \minus{}0.01 & \minus{}9.58 & \minus{}10.41 & \minus{}10.75 \\
    \cmidrule{2-9}
  & $\Delta(\theta) / \sigma(\theta)$ ($\Lambda$CDM) &
     0.53& \minus{}0.55& \minus{}0.41&     0.56&    \minus{}0.74&  ---  &  ---  \\
   & $\Delta(\theta) / \sigma(\theta)$ ($\wzero\wa$CDM) &   \minus{}0.03& 0.03& \minus{}0.44&    \minus{}0.01&    \minus{}0.19&    0.03& \minus{}0.12\\
    \midrule
     \midrule
\multirow{2}{*}{MCMC~~}  &  $1 - \sigma_\magn / \sigma_\text{no \magn}$ ($\Lambda$CDM) (\%)&    7.61 & \minus{3.53} & 0.0 & 7.75 & \minus{3.23} & --- & ---
    \\
    \cmidrule{2-9}
  & $\Delta(\theta) / \sigma(\theta)$ ($\Lambda$CDM) &
     0.71  & \minus{}0.66 & \minus{}0.36 & 0.72 & \minus{}0.81 & --- & --- \\
        \bottomrule
\end{tabular}
}
\end{table*}

As lensing magnification contains additional independent information, we expect the constraints to improve slightly when including it. In \Cref{tab:fisher_const_lensing_no_lensing} we report the improvement in constraints on cosmological parameters when magnification is included in the analysis. For a $\Lambda$CDM model, the impact of magnification on the reduction of the errorbars is $\lesssim 5\%$ for all cosmological parameters.
In the dynamical dark energy model $\wzero\wa$CDM, the impact of magnification is slightly larger, as the $1\sigma$ errorbars in $\wzero$ and $\wa$ are reduced by about $10\%$. {Nevertheless, the improvement due to magnification in the constraining power decreases for other parameters.}
It is important to note that in this test we are assuming the values of the local count slope to be exactly known.
While it is in principle possible to estimate $s(z)$ independently from the cosmological analysis, from the slope of the luminosity distribution of the galaxy sample, this measurement will be affected by several systematics, see for example~\citet{Hildebrandt:2015kcb}.
Therefore, we also consider a more pessimistic scenario where we assume no prior knowledge on the local count slopes and thus we marginalize over the values of $s$ in each redshift bins.
In \Cref{tab:fisher_const_lensing_no_lensing} we also compare the constraints on cosmological parameters for a model that neglects magnification, and a model that includes the effect, assuming no information on the local counts slope.
In this pessimistic setting, the constraints obtained when magnification is included are {worse} than the ones obtained when magnification is neglected.
This reduction in constraining power {as compared to a model without magnification} is up to 6\% for $\Lambda$CDM and becomes up to $10\%$ for the $\wzero\wa$CDM parametrization.
This is due to the fact that magnification does not contribute very significantly to the cosmological information that can be extracted from the 2-point correlation function, while the extra nuisance parameters introduced in this model slightly increase the degeneracy between the other parameters included in the analysis.
{We also note that, in the dynamical dark energy model, magnification mostly affects the cosmological constraints of $\sigma_8$, $\wzero$ and $\wa$, while the remaining model parameters are substantially unaffected.}
Nevertheless, the impact of magnification on the constraints of cosmological parameters is small; including it leads to changes of at most $\pm 10\%$ in the errorbars of parameters within the full-shape analysis.

\begin{figure*}
  \centering
  \includegraphics[width=\linewidth]{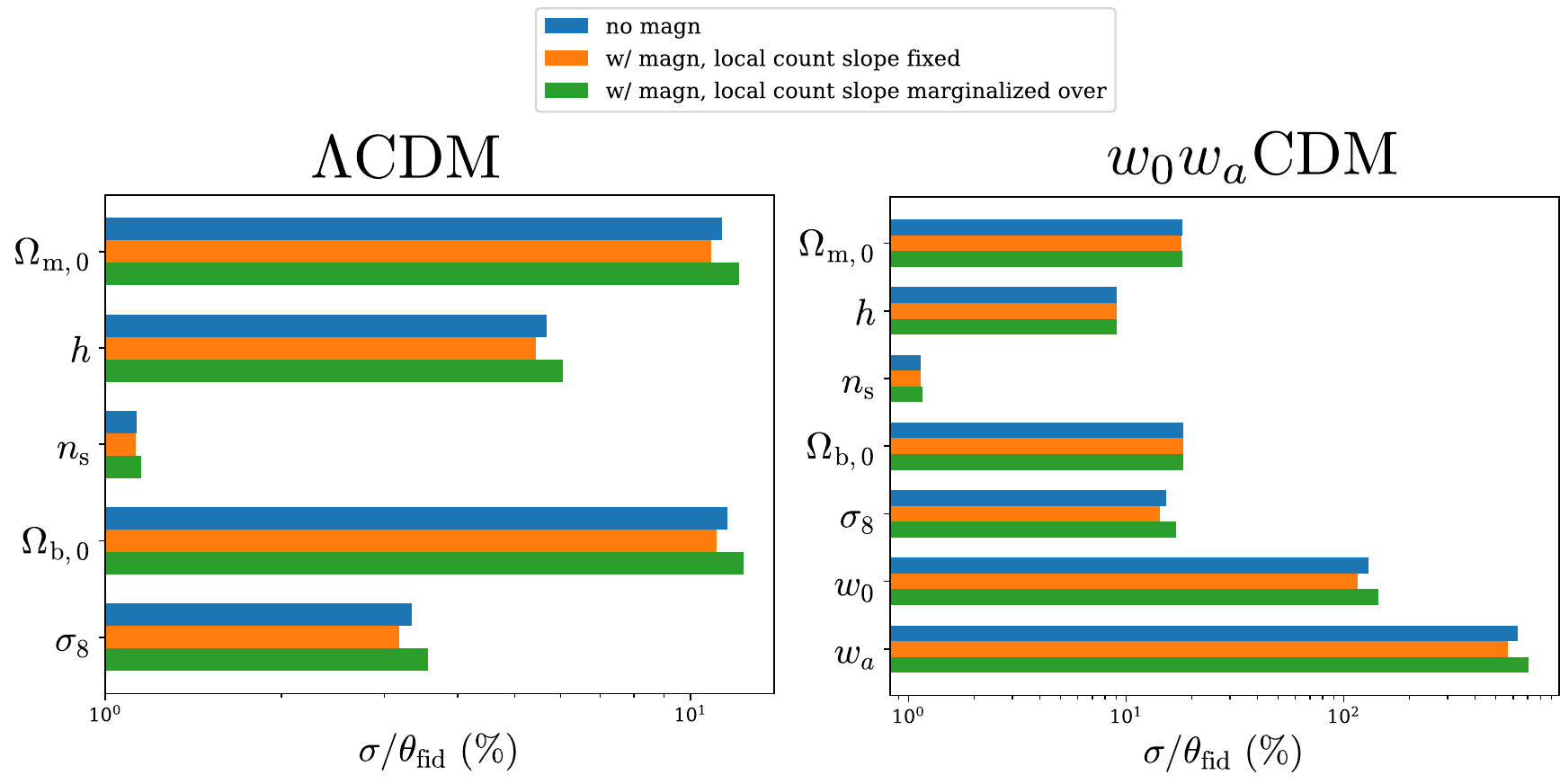}
    \caption{Comparison of {$68\%$ C.L.'s} obtained from the Fisher analysis for $\Lambda$CDM (left) and $\wzero\wa$CDM (right) with no lensing magnification (blue), with lensing magnification and the local count slope fixed (orange) in each redshift bin, and with magnification and local count slope marginalized (green).
    {Note that for $\wa$, we do not divide by the fiducial as it is zero, and instead we show the absolute error.}
    For the corresponding data, see \Cref{tab:fisher_const_lensing_no_lensing}.
    }
  \label{fig:fisher_lensing_wcdm_mag_bias}
\end{figure*}

\begin{figure*}
  \includegraphics[width=\linewidth]{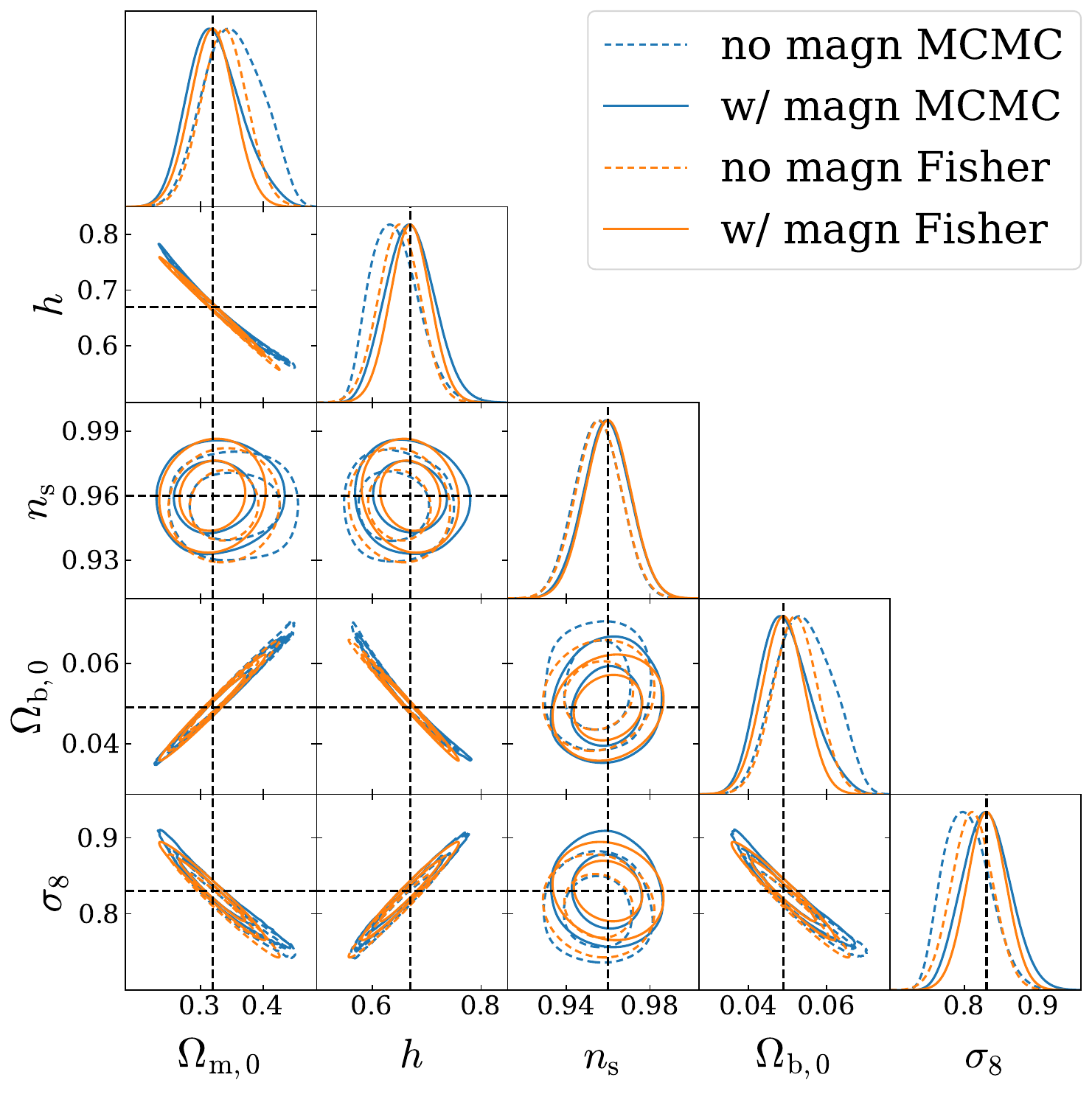}
    \caption{The {68\% (inner) and 95\% (outer) 2D confidence regions} and 1D posteriors for $\Lambda$CDM with marginalization over galaxy biases, using MCMC (blue) and the Fisher analysis (orange), with only contributions from just standard terms (dashed) as well as standard terms + lensing magnification (solid).
    The black dashed lines denote fiducial values of the cosmological parameters.}
  \label{fig:fisher_vs_mcmc_lcdm_big_omega}
\end{figure*}

In \Cref{fig:fisher_lensing_wcdm_mag_bias} we show a visual comparison of the constraints for the three cases discussed above.
We have validated these results by running an MCMC analysis for the $\Lambda$CDM model, including lensing
magnification in the analysis. A direct comparison of the contour plots for the Fisher and MCMC methods can be found in \Cref{fig:fisher_vs_mcmc_lcdm_big_omega} {and \Cref{tab:fisher_const_lensing_no_lensing}}. The Fisher results are accurate at the $20\%$ level (see \Cref{tab:fisher_extra} and \Cref{tab:mcmc_lcdm_big_small} in \Cref{app:additional} for the actual values).
{By comparing the results from the Fisher matrix and MCMC analyses in \Cref{fig:fisher_vs_mcmc_lcdm_big_omega}, it is easy to grasp the impact of the non-Gaussianity of the posterior. Such a non-Gaussianity, especially in the form of a skewness of the distribution, leads to a slight variation of the constraining power w/ and w/o magnification. In particular, this can be appreciated by comparing the last row of \Cref{tab:mcmc_lcdm_big_small} to the third one (\lcdm\ only). which also causes some changes in sign in some of the constraining power variation (e.g., first and second to last lines of \Cref{tab:fisher_const_lensing_no_lensing}): the change in the $68\%$ C.L.'s on those parameters switches sign. Note that all of these variations are very small---percentage level---and do not affect our main conclusions.}

We also investigate the effect of magnification on the accuracy of best-fit estimation of cosmological parameters. To accomplish this, we employ two distinct techniques, namely, Fisher analysis and MCMC analysis. In the Fisher analysis, we compute the biases in the best-fit estimates by employing \Cref{eq:shift}. In the MCMC analysis, we generate synthetic data based on a model that takes magnification into account and fit the data using two theoretical predictions: one that includes magnification, and the other that neglects it.
The differences in the best-fit parameters in these two cases provide the shifts induced by ignoring magnification in our modelling.
In \Cref{tab:fisher_const_lensing_no_lensing} (bottom block), we report the values of the shifts obtained with the Fisher analysis, for the $\Lambda$CDM and $\wzero\wa$CDM parametrizations.
In both cases, we find shifts below $1\sigma$.  For the  $\Lambda$CDM analysis, the best-fit estimate is biased at the level of  $\sim$0.5--0.7$\sigma$ for all cosmological parameters.
The impact is less relevant in the $\wzero\wa$CDM model, mainly due to the worse constraints on cosmological parameters.
The largest shifts in this case are found for $\ns$ ($\sim0.5\sigma$) and $\sigma_8$ ($\sim0.2\sigma$).
In \Cref{tab:fisher_const_lensing_no_lensing} (bottom block) we also report the MCMC result, generated only for $\Lambda$CDM, where the shifts are larger.
We find that while the shift found in the MCMC analysis is in most cases slightly larger, Fisher and MCMC forecast give consistent values of the shifts, both in terms of amplitude and direction. This provides an important check of the validity of the Fisher analysis.

One may wonder if shifts of less than a $1\sigma$ are something we should worry about. This means after all that the shifts are hidden in the uncertainty of the measurements. The goal of \Euclid is however to achieve an analysis where the sum of \emph{all} systematic effects is below 0.3$\sigma$. In this context, our analysis shows that including magnification in the modelling is necessary.

\begin{figure*}
  \includegraphics[width=\linewidth]{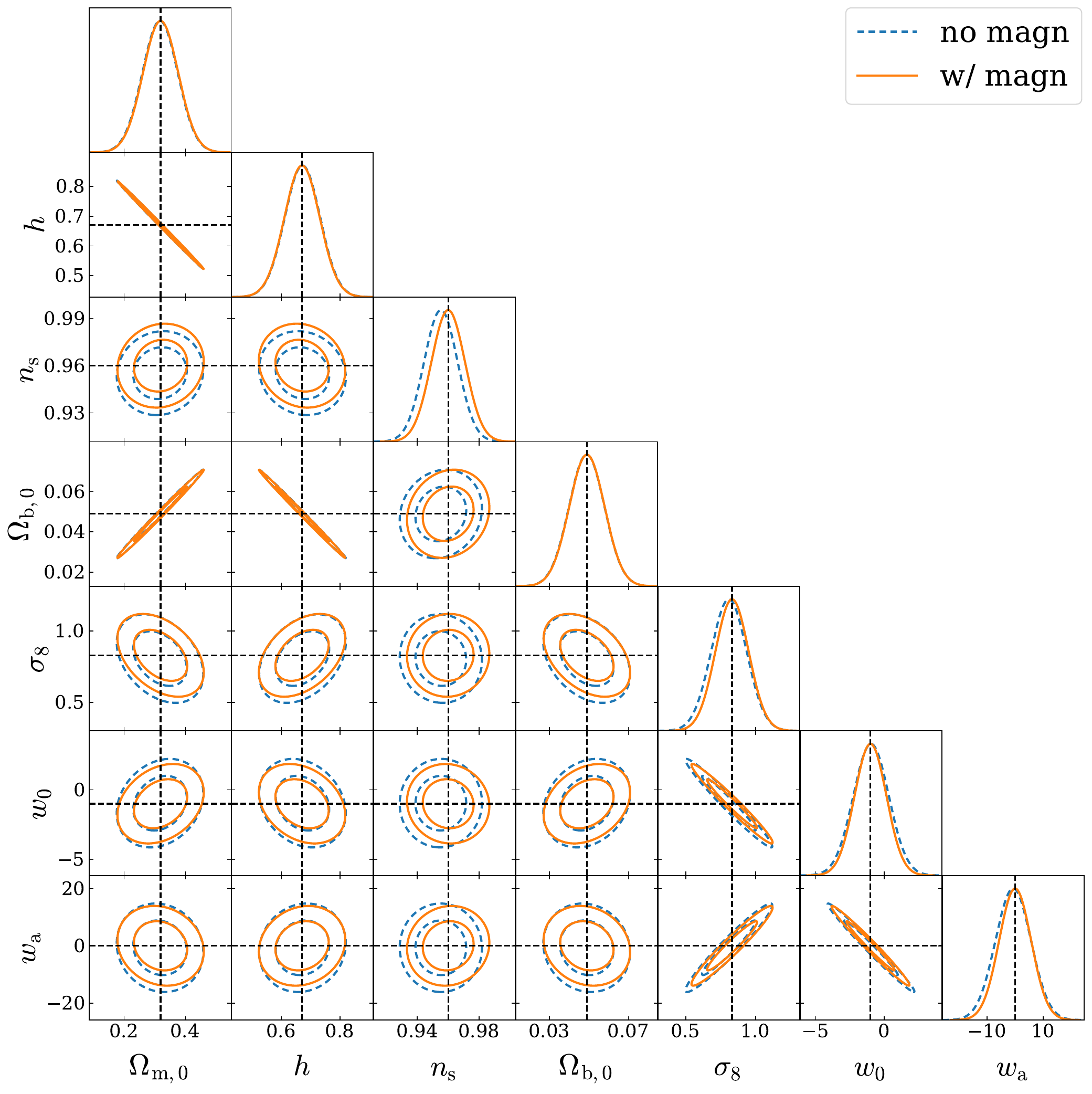}
    \caption{The {68\% (inner) and 95\% (outer) 2D confidence regions} and 1D posteriors for $\wzero\wa$CDM with marginalization over galaxy biases for the Fisher analysis, with contributions from standard terms (dashed) as well as standard terms + lensing magnification (solid). Black dashed lines denote the fiducial values. For corresponding values of the constraints and shift, refer to \Cref{tab:fisher_const_lensing_no_lensing}.}
  \label{fig:contour_wcdm_no_lensing_vs_lensing}
\end{figure*}

\begin{figure*}
  \includegraphics[width=\linewidth]{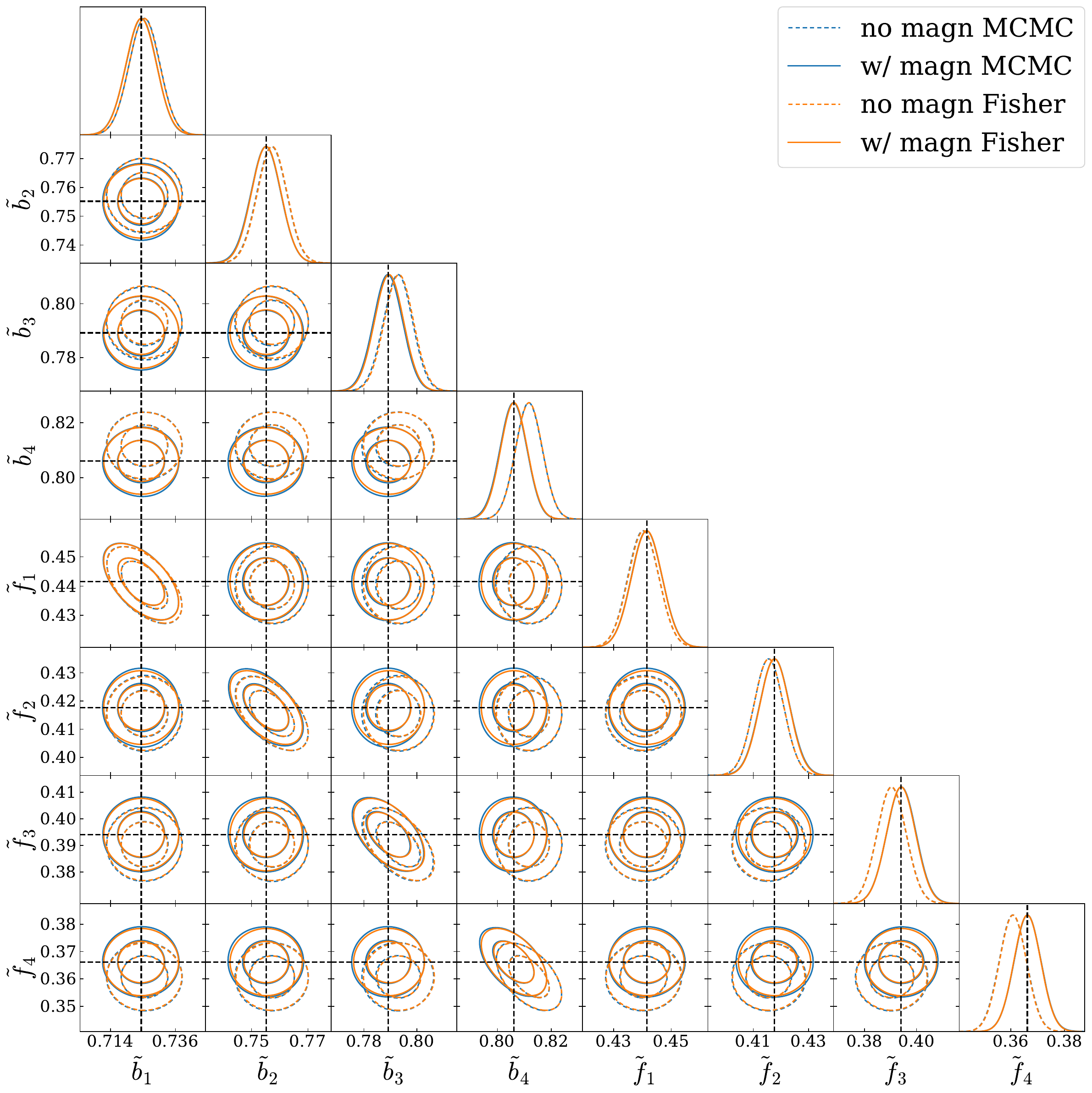}
    \caption{The {68\% (inner) and 95\% (outer) 2D confidence regions} and 1D posteriors for the $\{\tilde{f}, \tilde{b}\}$ parametrization, using MCMC (blue) and the Fisher analysis (orange), with only contributions from just standard terms (dashed) as well as standard terms + lensing magnification (solid). Black dashed lines denote the fiducial values. For corresponding values of the constraints and shift, refer to \Cref{tab:fisher_fb}.}
  \label{fig:fb_lensing_mcmc}
\end{figure*}

\begin{table*}[ht]
\centering
\caption{
    The constraints and shift for the $\tilde{f},\tilde{b}$ parametrization (\textit{top}: Fisher analysis; \textit{bottom}: MCMC analysis).
    Note that we model the magnification as being parametrization independent (i.e., a constant), hence the constraints obtained with magnification would be identical.
    \label{tab:fisher_fb}
}
\begin{tabular}{r r  r  r  r  r  r  r  r}
    \toprule
\backslashbox{quantity}{$\theta$}    &    $\tilde{b}_1$ & $\tilde{b}_2$ & $\tilde{b}_3$ & $\tilde{b}_4$ & $\tilde{f}_1$ & $\tilde{f}_2$ & $\tilde{f}_3$ & $\tilde{f}_4$ \\
    \midrule

    $\sigma(\theta)$ &       0.0052 &        0.0052 &        0.0055 &        0.0049 &        0.0053 &        0.0053 &        0.0056 &        0.0050 \\
  Fisher~~~~   $\sigma(\theta) / \theta(\%)$ &           0.71&        0.69&        0.69&        0.61&        1.20&        1.27&        1.41&        1.37\\
  $\Delta(\theta) / \sigma(\theta)$ &           0.21&        0.40&        0.68&        1.13&       \minus{}0.21&       \minus{}0.39&       \minus{}0.66&       \minus{}1.07\\
    \midrule
    \midrule
    $\sigma(\theta)$ &       0.0052 &        0.0053 &        0.0055 &        0.0050 &        0.0053 &        0.0054 &        0.0056 &        0.0051 \\
 MCMC~~~~  $\sigma(\theta) / \theta(\%)$ &           0.72&        0.70&        0.70&        0.62&        1.20&        1.29&        1.42&        1.39\\
    $\Delta(\theta) / \sigma(\theta)$ &           0.21&        0.40&        0.67&        1.16&       \minus{0.23}&       \minus{0.39}&       \minus{0.66}&       \minus{1.08}\\
    \bottomrule
\end{tabular}
\end{table*}
\subsection{Estimation of the growth rate}
As is clear from \Cref{eq:mult_z*}, the quadrupole and hexadecapole of the correlation function are most sensitive to the growth factor $f(z)$. Assuming that $\mu_4(d, z_*)$ is given by early Universe parameters determined by CMB observations, one might even use the hexadecapole alone to determine $\tilde f(z)$.
In practice, however, since the quadrupole and the monopole are much larger and correspondingly measured with much better precision, we use the combined multipoles of the correlations function to estimate both the bias and the growth factor together.
In this estimation we assume that the standard cosmological parameters are determined, e.g.,\ via CMB observations, and we only estimate the unknown functions $\tilde b(z) = \sigma_8(z) \,b(z)$ and $\tilde f(z) = \sigma_8(z) \, f(z)$.
Within GR we expect $f(z) \approx \Om_\text{m}(z)^{0.56}$, see \Cref{A:growth} for the exact expression.

We investigate the effect of lensing magnification on the estimation of the growth rate by fitting the full correlation function including lensing magnification with a model that does not include it.
We follow a similar approach as described in~\citet{Jelic-Cizmek:2020pkh, Breton:2021wiv}.
In each of the four redshift bins, we vary both the growth rate $\tilde f(z_i)=\tilde f_i$ and the bias $\tilde b(z_i)=\tilde b_i$.
The statistical error in the bias estimated via the Fisher analysis and via an MCMC study is typically of the order of $\sigma(\tilde b_i)\sim 0.7$\% while the error in the growth factor is of the order of $\sigma(\tilde f_i)\sim 1.2$\% to $1.4$\%, see \Cref{tab:fisher_fb}.
Note that in this case, adding the magnification in the model would not improve the measurement of $\tilde f_i$, since the magnification does not depend on these parameters. From \Cref{tab:fisher_fb}, we see that neglecting lensing magnification in the modelling shifts the best fit values of $\tilde b_i$ and $\tilde f_i$ by up to one standard deviation in the highest redshift bin, which is most strongly affected by magnification. But already in
bins two and three, neglecting lensing magnification leads to a systematic shift of more than 0.3$\sigma$, i.e.,\ above the target of \Euclid. Comparing the results from the Fisher analysis with those of the MCMC analysis (see \Cref{tab:fisher_fb} and \Cref{fig:fb_lensing_mcmc}), we find excellent agreement, both for the predicted constraints and for the shifts, even in the case where the shifts are larger than $1\sigma$.
This is on one hand due to the fact that the posteriors are very close to Gaussian, as can be seen from \Cref{fig:fb_lensing_mcmc}, and on the other hand, the derivatives of the signal with respect to $\tilde b_i$ and $\tilde f_i$ (which are used in the Fisher analysis) are trivial, since these parameters are constant coefficients (in each bin) in front of the scale-dependent functions $\mu_\ell(d,\bar{z})$, see \Cref{eq:mult_z*}.

From \Cref{fig:comparison_lensing_vs_no_lensing}, we see that the contribution from lensing magnification to all multipoles is positive. This is true at all redshifts, since $5s_i-2$ is positive in all bins (see \Cref{tab:gc_nofz}). As a consequence, lensing magnification increases the amplitude of the monopole and of the hexadecapole (that is positive) but it reduces the amplitude of the quadrupole (which is negative). Since the constraints come mainly from the monopole and the quadrupole, neglecting lensing magnification in the modelling means therefore that $\tilde b_i$ and $\tilde f_i$ are shifted in such a way that $\tilde b_i^2+2/3\tilde b_i\tilde f_i+\tilde f_i^2/5$ increases, while $4/3\tilde b_i\tilde f_i+4/7\tilde f_i^2$ decreases. This is best achieved by having a negative shift in $\tilde f_i$ and a positive shift in $\tilde b_i$.
Note that with these shifted values the hexadecapole will not be well fitted, because it would require an increase in $\tilde f_i$. But since its signal-to-noise ratio is significantly smaller than that of the monopole and the quadrupole, it does not have a significant impact on the analysis.

Such a systematic error in the analysis can certainly not be tolerated. Since the aim of the growth rate analysis is to test the theory of gravity, shifts of more than 1$\sigma$ in the growth rate would be wrongly interpreted as a detection of modified gravity. However, including lensing magnification in the analysis requires a model, which is exactly what we want to avoid in the growth rate analysis. In the case of the $\Lambda$CDM and $w_0w_a$CDM analyses, the problem is less severe since magnification can be modelled together with density and RSD. However, including it would significantly enhance the complexity of the computation and slow down the data analysis, especially for parameter estimation using MCMC methods. Below we propose a method to resolve these problems.

\subsection{A model for magnification as cosmology-independent systematic effect}
Since lensing magnification is a subdominant effect, we can include it in the modelling as a contamination, which does not encode any cosmological information, but that we can model sufficiently well.
More precisely, we pre-compute the lensing magnification with fixed cosmological parameters, e.g.,\ determined via CMB experiments and only vary the contributions from density and redshift space distortions in our analysis. {We do this in order to remove the bias of cosmological parameters which neglecting magnification can induce. On the other hand, this means that we lose the additional constraining power from lensing which, however, is not very significant.}
This `template method', which uses a fiducial template for the lensing magnification, has also been proposed in~\citet{Martinelli:2021ahc}.
Here we test it both on the $\Lambda$CDM and $w_0w_a$CDM analyses (where it is useful to reduce the computational costs) and on the growth rate analysis.
On the growth rate analysis, the template method allows us to preserve the model-independence of the method.
As explained before, in the growth rate analysis, the early time cosmology enters at the redshift $z_*$, before acceleration has started, and it is determined by CMB measurements.
The late time evolution is then \emph{fully} encoded in the parameters $\tilde f_i$ and $\tilde b_i$.
Any deviations in the laws of gravity would appear as a change in these parameters. Adding the lensing magnification to the signal would spoil the model-independence of the method, since this contribution cannot be easily written in a model-independent way.\footnote{As shown in~\citet{Tutusaus:2022cab}, the density-magnification term can be written in a way that does not depend on the late-time model, but the magnification-magnification is more involved due to the integral over the line-of-sight.} This would mean that part of the signal is modelled with $\tilde f_i$ and $\tilde b_i$, while the other part is modelled in a specific model, e.g.,\ in $\Lambda$CDM. We would then have a mix of parameters, some independent of the theory of gravity, and others specific to $\Lambda$CDM.
The template method circumvents this problem, by assuming that the lensing magnification is a fixed contribution, independent of cosmological parameters.
Of course this is not correct, but we show that the mistake that we make by doing this assumption does not introduce any significant shifts in the measurements of the variables $\tilde f_i$ and $\tilde b_i$ which we want to constrain with this method.
To test this, we include the lensing magnification in the model using \emph{wrong} cosmological parameters (since in practice we do not know the theory of gravity, nor the value of the true cosmological parameters).
More precisely, we use cosmological parameters that are one standard deviation below or above our fiducial values.
The $\pm 1\sigma$ values are reported in \Cref{tab:fid_template}.
We then compute the shifts in the cosmological parameters induced by the fact that the template for the lensing magnification is wrong by $\pm1\,\sigma$.

\begingroup
\setlength{\tabcolsep}{0.3em}
\begin{table}[ht!]
    \caption{Values of the cosmological parameters used for the template method.
    We consider 2 cosmologies, one $+1\sigma$ away from the fiducial (top), and one $-1\sigma$ away from the fiducial (bottom).
    Any other cosmological parameter not mentioned below is assumed to take its fiducial value (given in \Cref{tab:fid}).
    }
\begin{center}
\adjustbox{max width=\columnwidth}{
    \begin{tabular}{cccccccccc}
        \toprule
        case & $\Omm$ & $\Omb$ & $\sigma_8$ & $\ns$ & $h$ & $b_1$ & $b_2$ & $b_3$ & $b_4$ \\
        \midrule
        $+1\sigma$ & 0.354 & 0.054 & 0.856 & 0.971 & 0.706 & 1.488 & 1.683 & 1.899 & 2.122 \\
        $-1\sigma$ & 0.284 & 0.044 & 0.804 & 0.949 & 0.634 & 1.395 & 1.590 & 1.806 & 2.028 \\
        \bottomrule
    \end{tabular}}
\end{center}
\label{tab:fid_template}
\end{table}
\endgroup

\begin{table*}[ht]
    \centering
    \caption{
        The shift in $\Lambda$CDM (top) and $\wzero\wa$CDM (bottom) parameters obtained from a Fisher analysis using the template method, with parameters $+1\sigma$ (upper) and $-1\sigma$ (lower) away from the fiducial cosmology.
        To be conservative, we offset all of the galaxy biases according to $\max\{\sigma(b_i)\}$, $i \in \{1, \ldots, N\}$, where $N$ is the number of redshift bins in the survey (Fisher analysis).
    }
    \label{tab:fisher_lcdm_template}
\begin{tabular}{r r  r  r  r  r  r  r  r  r r}
    \toprule
    &    &     $\Omm$ &  $h$   &  $\ns$  & $\Omb$ & $\sigma_8$ & $\wzero$ & $\wa$\\
    \midrule
    \multirow{4}{*}{{$\Lambda$CDM}}~~~~    &    $\Delta(\theta)$                     &      \minus{}0.0018 & 0.0018 & \minus{}0.0004 &    \minus{}0.0003 &      0.002  & --- & --- \\
    &    $\Delta(\theta) / \sigma(\theta)$    &  \minus{}0.05& 0.05& \minus{}0.03&    \minus{}0.05&      0.07 & --- & --- \\
    \cmidrule{2-9}
    &    $\Delta(\theta)$                   &        0.0020 & \minus{}0.0020 & 0.0004 &     0.0003 &    \minus{}0.0021  & --- & --- \\
    &    $\Delta(\theta) / \sigma(\theta)$  &        0.05& \minus{}0.05& 0.03&     0.05&    \minus{}0.08 & --- & --- \\
    \midrule
    \midrule
\multirow{4}{*}{{$\wzero\wa$CDM}}~~~~    &    $\Delta(\theta)$                     &    \minus{}0.0002 & 0.0002 & \minus{}0.0003 &     0.0000 &     0.0019 & 0.0002 & 0.0363 \\
    &    $\Delta(\theta) / \sigma(\theta)$      &  \minus{}0.0038 & 0.0027 & \minus{}0.0313 &    \minus{}0.0008 &     0.0161 & 0.0002 & 0.0064 \\
    \cmidrule{2-9}
    &    $\Delta(\theta)$ &        0.0002 & \minus{}0.0002 & 0.0004 &     0.0000 &    \minus{}0.0022 & 0.0007 & \minus{}0.0448 \\
    &    $\Delta(\theta) / \sigma(\theta)$ &        0.0035 & \minus{}0.0025 & 0.0342 &     0.0006 &    \minus{}0.0182 & 0.0006 & \minus{}0.0079 \\
\bottomrule
\end{tabular}
\end{table*}

\begin{table*}[ht]
    \centering
    \caption{
        The shift in $\tilde{f}, \tilde{b}$ parameters using the template method, $+1\sigma$ (upper) and $-1\sigma$ (lower) away from the fiducial cosmology (\textit{top}: Fisher analysis; \textit{bottom}: MCMC analysis).
    }
    \label{tab:fisher_fb_template}
\begin{tabular}{r r r r  r  r  r  r  r  r  r  r  r  r}
    \toprule
      & &  $\tilde{b}_1$~~ & $\tilde{b}_2$~~ & $\tilde{b}_3$~~ & $\tilde{b}_4$~~ & $\tilde{f}_1$~~ & $\tilde{f}_2$~~ & $\tilde{f}_3$~~ & $\tilde{f}_4$~~ \\
\midrule
\parbox[t]{3mm}{\multirow{4}{*}{\rotatebox[origin=c]{90}{Fisher}}} ~~  &  $\Delta(\theta)$ & \minus{}0.0001 &       \minus{}0.0001 &       \minus{}0.0002 &       \minus{}0.0003 &        0.0001 &        0.0002 &        0.0004 &        0.0005 \\
  &  $\Delta(\theta)/\sigma(\theta)$    &              \minus{}0.0140 &       \minus{}0.0267 &       \minus{}0.0419 &       \minus{}0.0581 &        0.0214 &        0.0407 &        0.0654 &        0.0965 \\
    \cmidrule{2-10}
   &  $\Delta(\theta)$    &           0.0001 &        0.0002 &        0.0003 &        0.0004 &       \minus{}0.0001 &       \minus{}0.0002 &       \minus{}0.0004 &       \minus{}0.0005 \\
 & $\Delta(\theta)/\sigma(\theta)$      &           0.0166 &        0.0317 &        0.0505 &        0.0725 &       \minus{}0.0227 &       \minus{}0.0433 &       \minus{}0.0700 &       \minus{}0.1051 \\
\midrule
\midrule
\parbox[t]{3mm}{\multirow{4}{*}{\rotatebox[origin=c]{90}{MCMC}}} ~~
  &     $\Delta(\theta)$ & \minus{}0.0001 &       \minus{}0.0001 &       \minus{}0.0001 &       \minus{}0.0001 &        0.0001 &        0.0002 &        0.0003 &        0.0004 \\
  &  $\Delta(\theta)/\sigma(\theta)$    &              \minus{}0.0192 &       \minus{}0.0189 &       \minus{}0.0182 &       \minus{}0.02 &        0.0189 &        0.0370 &        0.0536 &        0.08 \\
    \cmidrule{2-10}
  &   $\Delta(\theta)$    &           0.0001 &        0.0003 &        0.0003 &        0.0005 &       0.0000 &       \minus{}0.0003 &       \minus{}0.0004 &       \minus{}0.0006 \\
 & $\Delta(\theta)/\sigma(\theta)$      &           0.0192 &        0.0566 &        0.0545 &        0.1000 &       0.0000  &       \minus{}0.0556 &       \minus{}0.0714 &       \minus{}0.1176 \\
       \bottomrule
\end{tabular}
\end{table*}

The shifts in the cosmological parameters for both $\Lambda$CDM and $\wzero\wa$CDM are given in \Cref{tab:fisher_lcdm_template}.
Comparing with \Cref{tab:fisher_const_lensing_no_lensing} we see that the shifts are very significantly reduced with the template method. In $\wzero\wa$CDM, the cosmological parameters never change by more than $0.04\sigma$.
In $\Lambda$CDM, $\sigma_8$ and the galaxy bias are shifted by $0.08\sigma$ to $0.13\sigma$ when using the template method.
Note, however that we did not vary $\wzero$ and $\wa$ for the template lensing magnification since these parameters are not well determined by {CMB data
used to obtain the template}.
The results for the growth rate analysis are presented in \Cref{tab:fisher_fb_template}.
Again, we see that the template method strongly reduces the shift, which, in the highest redshift bin goes down from $1\sigma$ (when magnification is fully neglected) to 0.1$\sigma$ with the template method. In the other three bins, the shifts are even smaller.

This template method, where magnification has to be computed only once, is therefore a very promising, inexpensive method to include lensing magnification in the analysis. Of course, once the best fit parameters are determined, one will want to include the magnification term with these parameters and run the analysis a second time in order to improve the fit. {This iterative method ensures us that the result is not sensitive to the initial best fit used to compute the magnification. This is particularly important in light of the current tensions between CMB and large-scale structure constraints.} Even though this method has the slight disadvantage that it does not use the information in the lensing magnification to constrain the cosmology, we believe that for a spectroscopic survey, for which magnification is weak, it is the simplest way to avoid the very significant biasing of the results which an analysis neglecting lensing magnification does generate, without significant numerical cost and with very minor loss of parameter {precision} (a few percent increase in the error bars).

\section{Conclusions}

\label{sec:con}
In this paper we studied the impact of lensing magnification on the spectroscopic survey of \Euclid.
Lensing magnification is commonly assumed to have a significantly lower impact on the spectroscopic analysis than the photometric one~\citep{Euclid:2021rez}, due to the fact that: 1) density fluctuations and RSD have higher amplitudes in a survey with spectroscopic resolution; 2) correlation between the different redshift bins are not taken into account in the spectroscopic analysis; and 3) the multipole expansion used in the spectroscopic analysis does remove part of the lensing signal (which is not fully captured by the first three multipoles). Despite this, we find that neglecting magnification leads to significant shifts in the cosmological parameters, of 0.2 to 0.7 standard deviations. Especially $\sigma_8$, but also $\Omm$ and $\Omb$ are shifted by more than half a standard deviation.

These shifts become even more significant when we consider the growth rates which we have fitted in an analysis that is independent of the late-time cosmological model.
If lensing magnification is neglected, the growth rate is shifted by more than one standard deviation at the highest redshift bin.

From these findings we conclude that the inclusion of lensing magnification in the data analysis of the spectroscopic survey of \Euclid is imperative.
In \Cref{sec:flatsky_expressions} we provide simplified expressions, based on the flat-sky approximation, for the contribution of magnification to the multipoles of the 2-point correlation function~\citep[see][for their derivation]{Jelic_Cizmek_2021}. While the cross-correlation of density and magnification can be computed very efficiently, the estimation of the magnification-magnification term is slowed down by an integral over the line-of-sight, which complicates the analysis of the $\Lambda$CDM and the $w_0w_a$CDM models. Even more importantly, the lensing magnification contribution cannot be easily written in a model-independent way. Consequently, including this contribution in the growth rate analysis would spoil the model-independence of the method. Since testing the laws of gravity with the growth rate analysis is one of the key goal of the spectroscopic analysis, this situation is problematic. Fortunately, we have proposed a method to solve the problem and reduce the shifts on the parameters to less than 0.1\,$\sigma$, while keeping the analysis independent of late-time cosmology. In this so-called `template method', lensing magnification is calculated in a model with fixed cosmological parameters and simply added to the standard terms. We have shown that if these fixed cosmological parameters deviate by 1$\sigma$ from the true underlying model, including this slightly wrong contribution from magnification leads to shifts in the inferred cosmological parameters by at most $0.1\sigma$. Of course one can then improve the analysis by iterating the process.

In our work we have compared our Fisher forecasts with a full MCMC analysis at several stages, and we have found that both methods
{provide consistent results for the parameter shifts even when they }are up to 1$\sigma$.

Note that unlike the previous \Euclid forecasts~\citepalias{Blanchard:2019oqi}, that are based on the power spectrum in Fourier space, our study presents a configuration-space analysis, based on the multipoles of the correlation function. Nevertheless, we expect that our conclusions similarly apply for a Fourier-space analysis. However, modelling the effects of magnification on the multipoles of the power spectrum is less straightforward.
{First}, the contribution of magnification depends on the power spectrum estimator and the survey window function.
{Second}, as demonstrated in~\citet{Castorina:2021xzs}, considering the contribution of magnification to the power spectrum multipoles necessitates prior estimation of the multipoles of the correlation function. Hence, our analysis is conducted directly in configuration space.
Even if the covariance matrix has more significant off-diagonal contributions in configuration space, this method has the advantage to be more direct and less survey dependent.

To conclude, we found that including magnification in the analysis does not significantly reduce the error bars on the inferred cosmological parameters. However, we will have to include it in our analysis of the data since otherwise we fit the data to the wrong physical model which will bias the inferred cosmological parameters.

\FloatBarrier

\begin{acknowledgements}
We acknowledge the use of the HPC cluster Baobab at the University of Geneva for conducting our numerical calculations.
GJC acknowledges support from the Swiss National Science Foundation (SNSF), professorship grant (No.~202671). FS, CB, RD and MK acknowledge support from the Swiss National Science Foundation Sinergia Grant CRSII5\underline{~}198674. CB acknowledges financial support from the European Research Council (ERC) under the European Union’s Horizon 2020 research and innovation program (Grant agreement No.~863929; project title ``Testing the law of gravity with novel large-scale structure observables''). LL is supported by a SNSF professorship grant (No.~202671).
PF acknowledges support from Ministerio de Ciencia e Innovacion, project PID2019-111317GB-C31, the European Research Executive Agency HORIZON-MSCA-2021-SE-01 Research and Innovation programme under the Marie Sk\l{}odowska-Curie grant agreement number 101086388 (LACEGAL). PF is also partially supported by the program Unidad de Excelencia María de Maeztu CEX2020-001058-M.
CV acknowledges an FPI grant from Ministerio de Ciencia e Innovacion, project PID2019-111317GB-C31.
{This work has made use of CosmoHub.
CosmoHub has been developed by the Port d'Informació Científica (PIC), maintained through a collaboration of the Institut de Física d'Altes Energies (IFAE) and the Centro de Investigaciones Energéticas, Medioambientales y Tecnológicas (CIEMAT) and the Institute of Space Sciences (CSIC \& IEEC), and was partially funded by the ``Plan Estatal de Investigación Científica y Técnica y de Innovación'' program of the Spanish government.}
\AckEC
\end{acknowledgements}

\bibliographystyle{aa}
\bibliography{biblio}

\appendix

\section{Code validation}
\label{sec:validation}
The analysis presented in this work is performed using the latest version of the code \coffe{}.
In order to make sure that the results of the analysis can be trusted, we first performed a code validation.
As a reference, we chose to use the well-established \cbl{} code from~\citet{cbl}, which can, among other outputs, compute the redshift-space multipoles of the 2-point correlation function.

\begin{figure}
  \includegraphics[width=\linewidth]{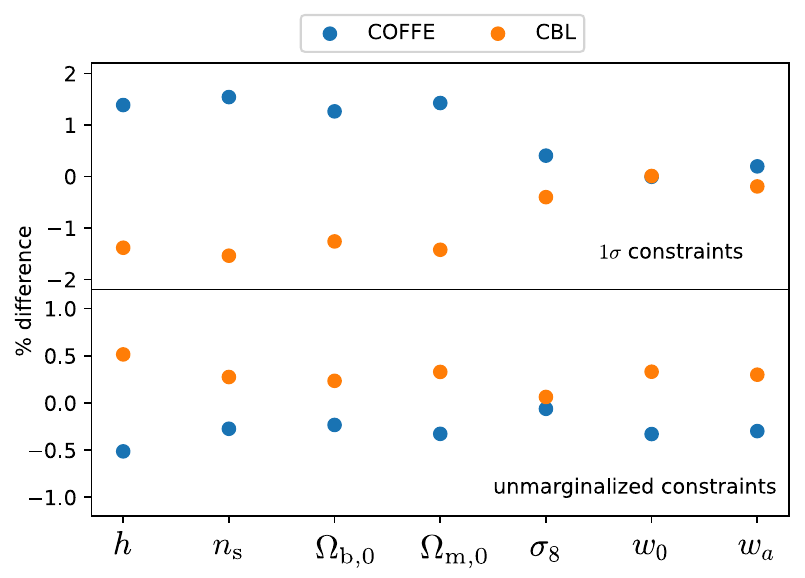}
  \caption{Percentage difference between \coffe{} and \cbl{} in the $1\sigma$ uncertainties (top panel) and unmarginalised constraints (bottom panel) for the spectroscopic sample of galaxy clustering.
This analysis includes 4 nuisance parameters for the galaxy bias that are marginalised over in the $1\sigma$ constraints.}
  \label{fig:code-valid-1}
\end{figure}

\begin{figure}
  \includegraphics[width=\linewidth]{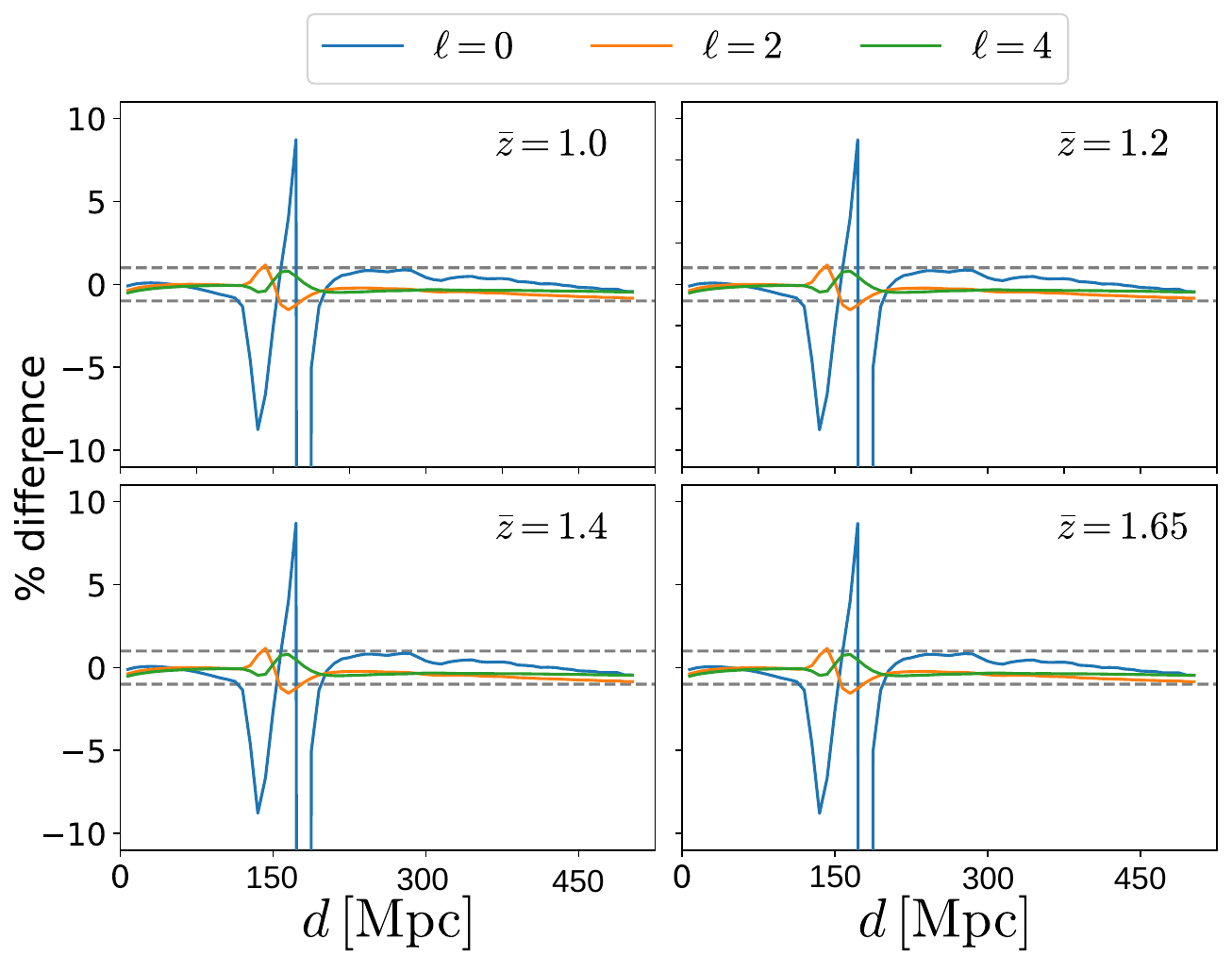}
  \caption{Percentage difference between \coffe{} and \cbl{} in the first 3 even multipoles of the 2PCF, for various redshifts.
The large ``jump'' of the monopole around {$r \sim 180\;\text{Mpc}$} is caused by its passage through zero.
The black dashed lines denote a 1\% threshold.}
  \label{fig:code-valid-0}
\end{figure}

\begin{figure}
  \includegraphics[width=\linewidth]{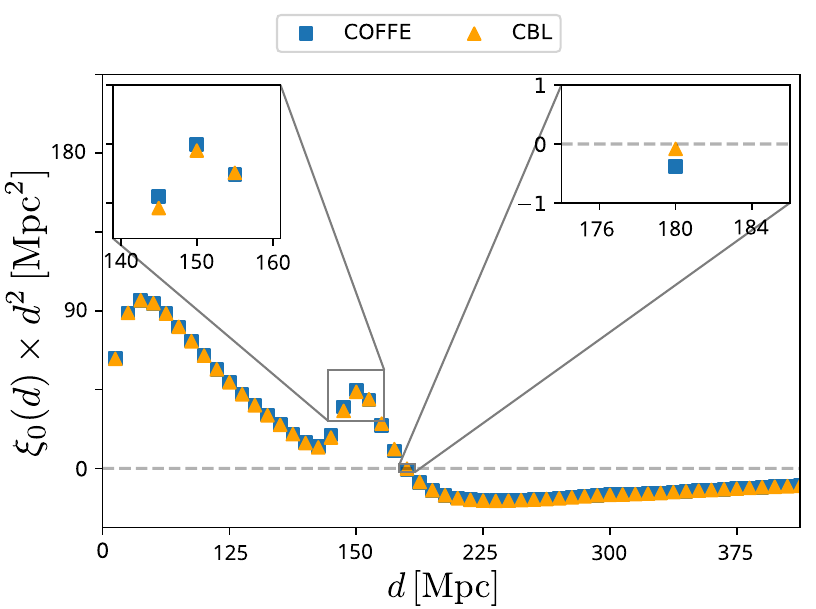}
  \caption{Comparison between the monopole from \coffe{} and \cbl{} at $\bar{z} = 1$.
The dashed line denotes the zero crossing.}
  \label{fig:code-valid-0b}
\end{figure}

The baseline settings used for this code comparison are the same as the ones adopted in~\citetalias{Blanchard:2019oqi} for the {$\text{GCsp}$ analysis}.
In summary:

\begin{itemize}
\item
The cosmological parameter space is $\theta = \{\Omm$,
$\Omb$, $\wzero$, $\wa$, $h$, $n_\text{s}$, $\sigma_8\}$, that
is, a flat cosmology with dynamical dark energy.
\item
The galaxy sample is split in 4 redshift bins, with a galaxy number density as
specified in \Cref{tab:gc_nofz}.
\item
We include the 4 galaxy bias parameters, one in each redshift bin, as nuisance parameters.
\item
As we are primarily interested in the validation using linear theory only, we set $r_\text{min} = 22\;$Mpc as the smallest separation in each redshift bin.
\end{itemize}
In \Cref{fig:code-valid-1}, we present the code comparison.
We show the percentage difference between the constraints obtained with the two codes and the mean values of the two results.
The top panel refers to $1\sigma$ marginalised constraints, while the bottom panel shows the comparison for the unmarginalised constraints.
The largest discrepancies between the two codes are $\sim 2\%$ for the $1\sigma$ errors and $\sim 1\%$ for the unmarginalised constraints.
We note that the outcome of the two codes has been compared for several intermediate steps, different settings, and different probe combinations, always leading to an excellent agreement.
In particular, we verified that using the covariance from either \coffe{} or \cbl{} when computing the constraints has no impact on the result;
we show a comparison of the two signals for the various redshifts in \Cref{fig:code-valid-0}. Even though the hexadecapoles show differences up to 10\% and larger in the vicinity of the baryon acoustic oscillation (BAO) peak, due to the small amplitude of this contribution, this is not relevant for parameter estimation.
In \Cref{fig:code-valid-0b} we show the monopole from both \coffe{} and \cbl{} at $z = 1$, with a close-up of some points of interest, notably, the BAO peak at $\sim 150\;$Mpc, and the zero crossing at $\sim 180\;$Mpc.

\begin{figure}
  \includegraphics[width=\linewidth]{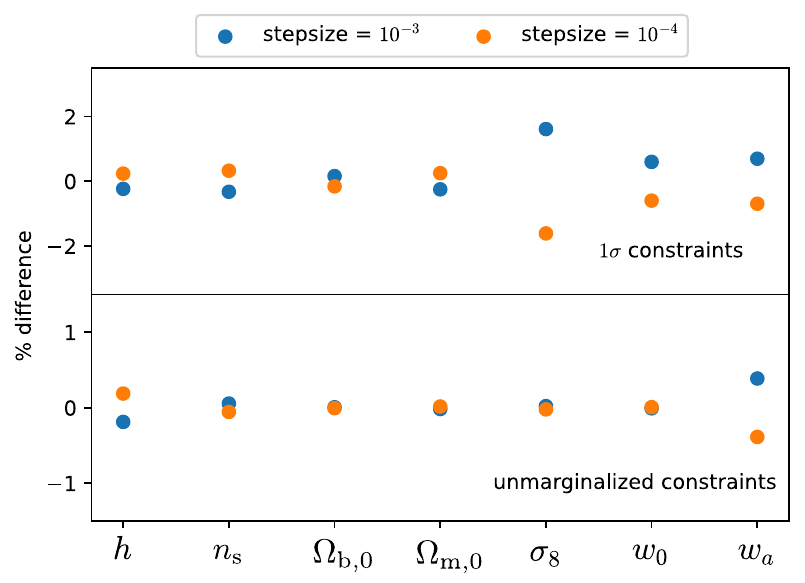}
  \caption{Percentage difference between stepsizes $10^{-3}$ and $10^{-4}$ in the $1\sigma$ uncertainties (top panel) and unmarginalised constraints (bottom panel) for the spectroscopic sample of galaxy clustering.}
  \label{fig:code-valid-2}
\end{figure}

In order to obtain the Fisher matrix, we need to compute derivatives of the multipoles of the 2PCF with respect to cosmological parameters.
As it is not possible to compute them analytically, we resort to the method of finite differences.
Since this method suffers from numerical instabilities, it is necessary to first find the optimal step size which is neither too large (causing the derivative to be too ``coarse''), nor too small (resulting in errors due to numerical underflow).
The final step size used ($10^{-3}$ for all parameters) has proven to be sufficiently accurate at the level of the obtained marginalized constraints for parameters of the \wcdm{} model. As shown in \Cref{fig:code-valid-2}, the difference of the inferred parameters for step sizes $10^{-3}$ and $10^{-4}$ is always below 2\% of the standard deviation.

\section{Fitting functions for $b(z)$ and $s(z)$}
For convenience, we provide a polynomial fit (obtained using the Levenberg-Marquardt algorithm) for both the galaxy bias and the local count slope as given below. We set
\begin{align}
b(z) &= \tilde{b}_0 + \tilde{b}_1 z + \tilde{b}_2 z^2 + \tilde{b}_3 z^3\,,\\
s(z) &= \tilde{s}_0 + \tilde{s}_1 z + \tilde{s}_2 z^2 + \tilde{s}_3 z^3\,,
\end{align}
with parameters:
\begin{equation}
\begin{array}{@{\hspace{5pt}}c@{\hspace{5pt}}c@{\hspace{5pt}}c@{\hspace{5pt}}c@{\hspace{5pt}}}
\tilde{b}_0 = 0.853, & \tilde{b}_1 = 0.040, & \tilde{b}_2 = 0.713,  & \tilde{b}_\fra{3} = -0.164, \\
\tilde{s}_0 = 1.231, & \tilde{s}_1 = -1.746, & \tilde{s}_2 = 1.810, & \tilde{s}_3 = -0.505.
\end{array}
\label{eq:fit_biases}
\end{equation}
In \Cref{f:b_and_s} we compare our best fit with the measurements of the Flagship simulation.
In our calculations we do not use these fits, but we present them here for convenience.
The Flagship specifics have been estimated for the survey binning as described in \Cref{sec:flagship_specs}, and therefore the fitting functions are adapted to this specific configuration.

\section{Binned covariance of 2PCF multipoles}
\label{ap:cov}

In this section we report the exact expression for the covariance adopted in this
work. We modified the implementation of the flat-sky, Gaussian covariance reported in~\citet{coffe} in order to include a binning of the Bessel functions.
The full expression of the covariance can be written
in the same form as~\citet{coffe}, that is

\begin{gather}
\begin{aligned}
\covariance\left[\xi^j_{\ell},\xi^k_{\ell'}\right] &= \frac{i^{\ell-\ell'}}{V}
\left[\frac{1}{\bar{\numberdensity}}
\bar{\mathcal{G}}_{\ell \ell'} (d_j, d_k, \bar{z} )\sum_\sigma c_\sigma
\begin{pmatrix}
\ell & \ell' & \sigma \\
0 & 0 & 0
\end{pmatrix}^2 \right.\\
& \left. +\bar{\mathcal{D}}_{\ell \ell'} (d_j, d_k, \bar{z}) \sum_\sigma \tilde{c}_\sigma
\begin{pmatrix}
\ell & \ell' & \sigma \\
0 & 0 & 0
\end{pmatrix}^2 \right. \\
& \left. + \frac{2\ell + 1}{2\pi \bar{\numberdensity}^2 d_j^2 L_p} \delta_{jk} \delta_{\ell \ell'}\,   \right],
\end{aligned}
\end{gather}
where the $\begin{pmatrix}
    \ell & \ell' & \sigma \\
    0 & 0 & 0
\end{pmatrix}$ denote Wigner 3j symbols, hence the sum over $\sigma$ goes from $|\ell-\ell'|$ to $\ell+\ell'$. The three terms in the sum respectively denote the cross-correlation between cosmic variance and Poisson noise, the cosmic variance autocorrelation, and the Poisson noise autocorrelation.
{Here $V$ denotes the comoving volume of the observed sample, $\numberdensity$ denotes the average comoving number density of sources, and $L_p$ denotes the pixelsize, that is, the minimum comoving distance we can resolve.}
Note that the coefficients $\{c_\sigma, \tilde{c}_\sigma\}$ depend only on redshift. Their exact expressions are reported in~\citet{coffe}, and we repeat them here for completeness.
\begingroup
\allowdisplaybreaks
\begin{align}
&c_0 =  b^2+\frac{2}{3}bf +\frac{f^2}{5} \,, \label{e:c0}\\
&c_2 =  \frac{4}{3} bf +\frac{4}{7} f^2 \,, \\
&c_4 = \frac{8}{35} f^2\,, \label{e:c4} \\
&\tilde  c_0 = c_0^2 +\frac{c_2^2}{5}+\frac{c_4^2}{9} \,, \label{e:c0tilde}\\
&\tilde  c_2 = \frac{2}{7}c_2 (7c_0+c_2) +\frac{4}{7} c_2 c_4+\frac{100}{693}c_4^2 \,, \\
&\tilde  c_4 = \frac{18}{35} c_2^2 +2 c_0 c_4 +\frac{40}{77} c_2 c_4 +\frac{162}{1001} c_4^2 \,,\\
&\tilde  c_6 = \frac{10}{99} c_4 (9 c_2 +2 c_4) \,,\\
&\tilde  c_8 = \frac{490}{1287} c_4^2 \,. \label{e:c8}
\end{align}
\endgroup
The main difference between the original implementation in \coffe{} and the covariance used in this analysis lies in the computation of $\bar{\mathcal{G}}_{\ell \ell'}$ and $\bar{\mathcal{D}}_{\ell \ell'}$.
Here these are estimated as integrals of the binned spherical Bessel functions
\begin{equation}
\bar{j}_\ell (k d_i) := \frac{4\pi}{V_{d_i}} \int_{d_i - L_p/2}^{d_i + L_p/2} \diff s \; s^2 j_\ell (k s),
\end{equation}
where $V_{d_i} = \frac{4\pi}{3} \left(d_{i, \text{max}}^3 -d_{i, \text{min}}^3 \right)$ is the volume of the distance bin around $d_i$.
Thus, we have
\begin{align}
  \frac{\bar{\mathcal{G}}_{\ell \ell'} (d_i, d_j, \bar{z} )}{(2\ell + 1)(2\ell' + 1)}
  &= \frac{2}{\pi^2}
  \int_0^\infty \diff k\,k^2 P(k, \bar{z})\,\bar{j}_\ell (k d_i) \bar{j}_{\ell'} (k d_j)\,, \notag\\
  \frac{\bar{\mathcal{D}}_{\ell \ell'} (d_i, d_j, \bar{z} )}{(2\ell + 1)(2\ell' + 1)}
  &= \frac{1}{\pi^2}
  \int_0^\infty \diff k\,k^2 P^2(k, \bar{z})\,\bar{j}_\ell (k d_i) \bar{j}_{\ell'} (k d_j)\,.\notag
\end{align}
In fact, it is shown in the literature that the covariance matrix is overestimated when this volume-average over the spherical Bessel functions is not applied, see~\citet{Grieb:2015bia}.

\begin{figure}
  \includegraphics[width=\linewidth]{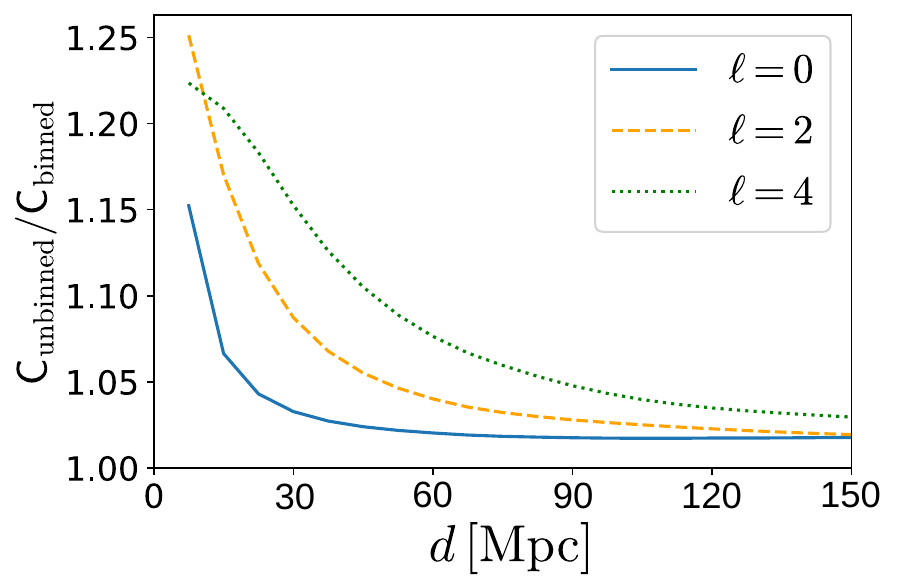}
  \caption{The diagonal entries of the ratio of the unbinned and the binned covariance, for various multipoles, as a function of comoving separation $d$, at the lowest redshift bin of \Euclid.}
  \label{fig:covariance_comparison}
\end{figure}

In \Cref{fig:covariance_comparison} we show the ratio of the diagonal entries of the unbinned and the binned covariance; as we can see, for low separations and large multipoles, the unbinned covariance can be larger than its binned counterpart by more than 20\%.

\section{An analytic expression for the growth rate}\label{A:growth}

\begin{figure}
\includegraphics[width=\linewidth]{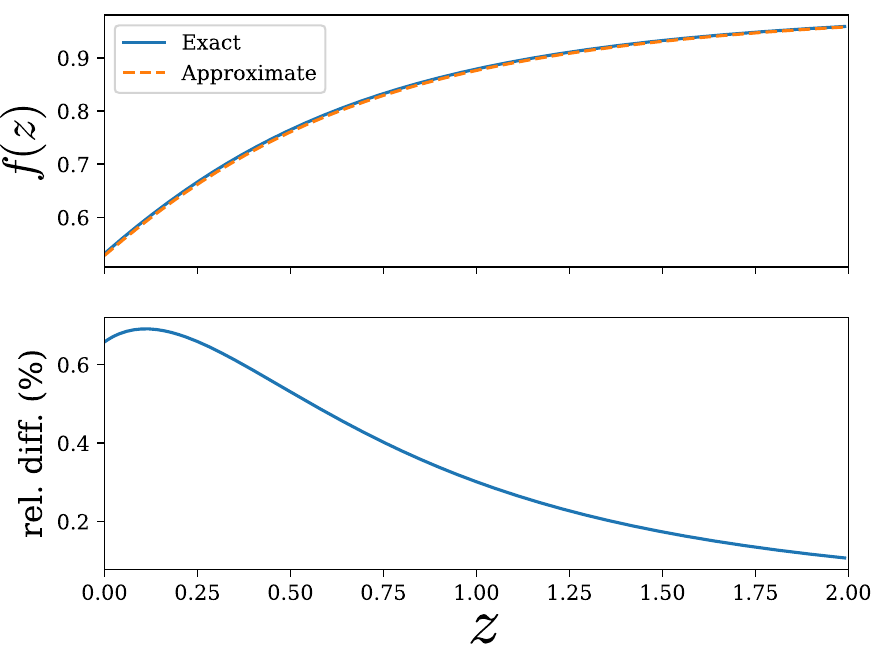}
\caption{\textit{Top}: the plot of the exact expression for the growth rate, given by \Cref{eq:f_exact} (blue solid curve), and the approximate expression, given by \Cref{eq:f_approx} (orange dashed curve).
\textit{Bottom}: the relative difference between them, in percentage points. In both cases we assume $\Omm=0.32$.}
\label{f:comparison_f}
\end{figure}

In \Cref{2pf} we introduced the growth rate $\ng{f(z):=\frac{\mathrm d\ln\delta}{\mathrm d\ln a}}\,$. Here we derive an analytical expression for $f(z)$ in linear perturbation theory and compare it to the commonly used expression $f(z)\approx \Om_\text{m}(z)^{0.56}$. The linear density fluctuation in comoving gauge is $\delta(z) \approx D_1(z)\, \delta(0) $. Following~\citet{durrer_2020}, we obtain:
\begin{align} \label{D}
\ng{f(z) =\frac{\mathrm d\ln D_1}{\mathrm d\ln a} = - (1+z)\frac{\mathrm d\ln D_1}{\mathrm d z}}\,,
\end{align}
where $D_1$ is the linear growth factor. In a $\Lambda$CDM universe, $D_1$ is {the growing mode solution of} the following equation:
\begin{align} \label{D1}
    \ddot{D_1}+ \HH \dot{D_1} = \frac{3}{2} \HH^2 \Om_\text{m}(a)D_1,
\end{align}
with
\begin{align}
    \Om_\text{m}(a)=\frac{\Omm \, a^{-3}}{\Omm \, a^{-3}+(1-\Omm)}.
\end{align}
We rewrite \Cref{D1} using a prime to indicate the derivative with respect to $\ln a$:
\begin{align}\label{e:D''}
    D_1''+ \left( 2 - \frac{3}{2} \Om_\text{m}(a) \right) D_1'  = \frac{3}{2} \Om_\text{m}(a)D_1,
\end{align}
The analytical solution of \Cref{e:D''} in terms of hypergeometric functions~\citep[see][]{Abra}, leads to the following expression for the growth rate:
\begin{align}
    f(z)=\frac{1}{2}\Om_\text{m}(z) \left[ \frac{5}{_2F_1 \left ( \frac{1}{3},1;\frac{11}{6};1-\frac{1}{\Om_\text{m}(z)} \right) } - 3 \right].
    \label{eq:f_exact}
\end{align}
In the literature \citep[see, e.g.,][]{durrer_2020, Linder_2007}, one often finds the following approximation for the growth rate:
\begin{align}
    f(z) \approx \Om^{0.56}_\text{m}(z)\,.
    \label{eq:f_approx}
\end{align}
A comparison between eqs. \Cref{eq:f_exact} and \Cref{eq:f_approx} is shown in \Cref{f:comparison_f}. The approximation is clearly excellent, leading to differences below 1\% at all redshifts.

\section{Expressions for the multipoles of the 2PCF}
\label{sec:expressions}
\subsection{Curved-sky expressions}
\label{sec:fullsky_expressions}

For completeness, below we state the relevant curved-sky contributions for the 2-point correlation function which were first derived in~\citet{Tansella_2018}.
{Note that, for brevity, in the below we use the coordinates $r_1 := r(z_1)$, $r_2 := r(z_2)$, and $\theta$, denoting the comoving distances at redshifts $z_1$ and $z_2$, and the angle at the observer, respectively.
Furthermore, we use $x_i := x(z_i)$.}

We use the following notation, for local terms $A$ and $B$:
\begin{gather}
\begin{aligned}
\xi^{AB}(d, \bar z, \mu)
&=
D_1(z_1)\,
D_1(z_2) \\
&\times \sum_{\ell,n}
\left(
X^n_\ell\big|_A
+
X^n_\ell\big|_{AB}
+
X^n_\ell\big|_{BA}
+
X^n_\ell\big|_B
\right)I_\ell^n(d),
\end{aligned}
\end{gather}
and the following for the integrated terms:
\begin{equation}
\xi^{AB}(d, \bar z, \mu)
=
Z\big|_A
+
Z\big|_{AB}
+
Z\big|_{BA}
+
Z\big|_B.
\end{equation}
\begingroup
\allowdisplaybreaks
\begin{align}
&X_0^0 \big|_\text{den} = b_1 b_2\,,
\label{eq:density}
\\
&X_0^0 \big|_\text{RSD} =  f_1 f_2   \frac{1+2\cos^2\theta}{15 }\,,
\label{eq:rsd1}
\\
&X_2^0 \big|_\text{RSD} = -\frac{f_1 f_2}{21}\left[1+11\cos^2\theta +\frac{18\cos\theta(\cos^2\theta-1){r_1}{r_2}}{d^2}\right] \,,\raisetag{1.5\baselineskip}
\label{eq:rsd2}
\\
&X_4^0 \big|_\text{RSD} = \frac{f_1 f_2}{35r^4} \big\{
    {4(3\cos^2\theta-1)({r_1}^4+{r_2}^4)} \nonumber \\
    &\qquad\qquad+ {{r_1}{r_2}} (3+\cos^2\theta)\big[ 3 (3+\cos^2\theta){r_1}{r_2} \label{eq:rsd3} \\
&\qquad\qquad -8({r_1}^2+{r_2}^2)\cos\theta \big] \big\} \nonumber\,,
\\
&X_0^0 \big|_\text{den-RSD} = \frac{b_1 f_2 }{3} \,,
\label{eq:den-rsd1}
\\
&X_2^0 \big|_\text{den-RSD} = -b_1 f_2  \left[\frac{2}{3} -(1-\cos^2\theta)\frac{{r_1}^2}{d^2}\right] \,,
\label{eq:den-rsd2}
\end{align}
{where above and below the indices 1 and 2 indicate that the corresponding quantities are evaluated at redshift $z_1$ respectively $z_2$ corresponding to the pair of voxels.}
\twocolumn[{
\begin{align}
&Z \big|_\text{den-\magn} = - \frac{3\Omm}{2} b_1 \frac{H_0^2}{c^2} \frac{2-5s_2}{{r_2}}D_1(z_1) \int\limits_0^{{r_2}} \mathrm{d} \la \frac{{r_2}-\la}{\la}\frac{D_1(\la)}{ a(\la)} \bigg[
  2\chi_1\la\cos\theta I^1_1(d) - \frac{{r_1}^2\la^2(1-\cos^2\theta)}{d^2} I^0_2(d)\bigg] \,,
\label{eq:den-len}
\end{align}
\begin{gather}
\begin{aligned}
&Z \big|_\magn
=
\frac{9 \Omm^2}{4}\frac{H_0^4}{c^4}\frac{(2-5s_1)(2-5s_2)}{{r_1}{r_2}} \int\limits_0^{{r_1}} \!\mathrm{d}\la \int\limits_0^{{r_2}} \!\mathrm{d}\la'
   \frac{({r_1}-\la)({r_2}-\la')}{\la \la'} \frac{D_1(\la)D_1(\la')}{a(\la)a(\la')} \bigg\{
\frac{2}{5} (\cos^2\theta-1) \la^2 \la'^2 I^0_0(d)  \\
& \qquad
+\frac{4 d^2 \cos(\theta) \la \la'}{3} I^2_0(d) +
\frac{4 \cos(\theta) \la \la' [d^2 +6 \cos(\theta) \la \la']}{15} I^1_1(d)
+
\frac{2(\cos^2\theta -1)\la^2\la'^2[2r^4 +3 \cos(\theta) d^2 \la \la']}{7 d^4} I^0_2(d)
\\
& \qquad
+\frac{2 \cos(\theta) \la \la' \left[2 d^4 +12\cos(\theta) d^2 \la\la' +15 (\cos^2\theta-1)\la^2\la'^2 \right]}{15 d^2} I^1_3(d)
\\
& \qquad
+\frac{(\cos^2\theta-1)\la^2\la'^2 \left[6 d^4 +30\cos(\theta) d^2\la\la' +35 (\cos^2\theta -1)\la^2\la'^2 \right]}{35d^4} I^0_4(d)\bigg\} \,,
\label{eq:len-len}
\end{aligned}
\end{gather}
\vspace{1cm}
}]
\endgroup
\noindent where
\begin{align}
    I^n_\ell(d) &= \frac{1}{2 \pi^2} \int_0^\infty \diff{} k \, k^2 \, P_{\delta\delta}(k, z = 0) \frac{j_\ell(k d)}{(k d)^n}.
\end{align}
Note that inside the integral, $d^2 = {r_2}^2 + \lambda^2 - 2 {r_2} \lambda \cos\theta$ in the case of density-magnification, and $d^2 = \lambda_1^2 + \lambda_2^2 - 2 \lambda_1 \lambda_2 \cos\theta$ in the case of magnification-magnification, while $\theta$ is the angle at the observer between the two lines of sight.
The result for the other cross-correlations can be obtained by performing the substitution $2 \leftrightarrow 1$.

\subsection{Flat-sky expressions}
\label{sec:flatsky_expressions}

The flat-sky expressions for the multipoles of the density-magnification and magnification-magnification terms implemented in \coffe{} are given by~\citep[see][]{Jelic_Cizmek_2021}
\begin{gather}
\begin{aligned}
\xi^\text{den-\magn}_{\ell}
(d, \bar z)
=&
-\frac{2\ell + 1}{2}
\frac{3}{8\pi}
\Omm{} \frac{H_0^2}{c^2}
D_1^2(\bar z)
(1 + \bar z)
d^2
\\
&\hspace{-1.5cm}\times
\left\{[2-5s_1(\bar z)]b_2(\bar z) +[2-5s_2(\bar z)]b_1(\bar z)\right\}
\\
&\hspace{-2cm}\times
\pi^\frac{3}{2}\
\frac{
2^\frac{5}{2}
}
{
2^\frac{\ell}{2}
}
\sum\limits_{k = 0}^{\left \lfloor \frac{\ell}{2} \right \rfloor}
\frac{(-1)^k}{2^k}
\begin{pmatrix} \ell \\ k \end{pmatrix}
\begin{pmatrix} 2\ell - 2k \\ \ell \end{pmatrix}
\left[\frac{\ell}{2} - k\right]!\,
I^{\ell / 2 - k + 3 / 2}_{\ell / 2 - k + 1 / 2} (d),
\label{eq:den_len_analytic}
\end{aligned}
\end{gather}

\begin{gather}
\begin{aligned}
\xi^\text{\magn-\magn}_{\ell}
(d, \bar z)
=&
C(\ell)
\frac{(3\Omm{} H_0^2)^2[2-5s_1(\bar z)][2-5s_2(\bar z)]}{8 \pi c^4 \, \bar{r}^2}
\\
&\hspace{-2cm}\times
\int_0^{\bar{r}}
\diff \lambda\,
\left[
\lambda(\bar{r}-\lambda)
\right]^2\,
D_1^2[z(\lambda)]\,
[1+z(\lambda)]^2\,
\mathcal{K}_\ell\left(\frac{\lambda}{\bar{r}} d\right), \raisetag{1.8\baselineskip}
\label{eq:lens_lens_analytic}
\end{aligned}
\end{gather}
where $\bar{z} = (z_1 + z_2) / 2$ is the mean redshift, $\bar{r} = r(\bar{z})$ is the comoving distance evaluated at $\bar{z}$, $\left \lfloor \cdot \right \rfloor$ denotes the floor function, and we defined
\begin{align}
    C(\ell)&=\frac{2 \ell + 1}{2}\frac{\ell!}{2^{\ell - 1} [(\ell/2)!]^2}, \\
    \mathcal{K}_\ell(d) &= 2 \pi^2 d \, I_\ell^1(d).
\end{align}
\section{Additional material}
\label{app:additional}

In this appendix, we include two additional tables that complement the content presented in \Cref{sec:results}. Specifically, in \Cref{tab:fisher_extra}, we provide supplementary information regarding the Fisher full-shape analysis for both the $\Lambda$CDM and $\wzero\wa$CDM models.
Furthermore, in \Cref{tab:mcmc_lcdm_big_small}, we present similar results for the MCMC analysis, focusing on the two parametrizations of the $\Lambda$CDM model: $\{\Omm, \Omb\}$ (baseline analysis) and $\{\omm, \omb\}$. The $\{\omm, \omb\}$ parametrization offers the advantage of faster convergence in the MCMC analysis, as the posterior distribution of the cosmological parameters becomes more Gaussian.
\begin{table*}[ht]
    \centering
    \caption{
        The constraints for a $\Lambda$CDM (top) and a $\wzero\wa$CDM (bottom) cosmology obtained from the Fisher forecast, without lensing magnification (upper) and with lensing magnification (middle), as well as the shift (lower).
        Note that the last row in the table uses the constraints from the case with lensing magnification.
        Also note that the percentage constraints for $\wa$ are undefined since the fiducial value is 0, and hence here we just show the result as if it had a fiducial value of 1 instead.
        \label{tab:fisher_extra}
    }
\begin{tabular}{r r  r  r  r  r  r  r  r  r}
    \toprule
    \backslashbox{quantity}{$\theta$}    &    $\Omm$ &  $h$  & $\ns$ & $\Omb$ & $\sigma_8$  & $\wzero$ & $\wa$\\
    \midrule
    $\sigma(\theta)$ &        0.0361 &  0.0380 &  0.0109 &     0.0057 &     0.0277  & --- & --- \\
    $\sigma(\theta) / \theta(\%)$  &       11.32&  5.67&  1.13&    11.56&     3.33 & --- & --- \\
    \midrule
    $\sigma(\theta)$ (L) &        0.0346 &  0.0364 &  0.0108 &     0.0054 &     0.0263  & --- & --- \\
    $\sigma(\theta) / \theta(\%)$ (L) &       10.85&  5.43&  1.12&    11.06&     3.17 & --- & --- \\
    \midrule
    $\Delta(\theta)$ &        0.0185 & \minus{}0.0199 & \minus{}0.0044 &     0.0031 &    \minus{}0.0195  & --- & --- \\
    $\Delta(\theta) / \sigma(\theta)$ &        0.53& \minus{}0.54& \minus{}0.41&     0.56&    \minus{}0.73 & --- & --- \\
    \midrule
   \midrule
    $\sigma(\theta)$     &         0.057 & 0.061 &  0.011 &     0.009 &     0.127 &    1.294 &  6.302 \\
    $\sigma(\theta) / \theta(\%)$      &   18.06& 9.03&  1.13&    18.30&    15.3 & 129.4 & 630.2\\
    \midrule
     $\sigma(\theta)$ (L)     &   0.057 & 0.060 &  0.011 &     0.009 &     0.119 &    1.162 &  5.678 \\
    $\sigma(\theta) / \theta(\%)$ (L)    &        18.01& 9.01&  1.13&    18.24&    14.29& 116.2 & 567.7\\
    \midrule
    $\Delta(\theta)$      &   \minus{}0.0019 & 0.0015 & \minus{}0.0048 &    \minus{}0.0001 &    \minus{}0.023 &    0.040 & \minus{}0.682 \\
    $\Delta(\theta) / \sigma(\theta)$     &   \minus{}0.033& 0.024& \minus{}0.437&    \minus{}0.011 &    \minus{}0.191 &    0.034 & \minus{}0.120 \\

        \bottomrule
\end{tabular}
\end{table*}

\begin{table*}[ht]
    \centering
    \caption{
        The constraints for a $\Lambda$CDM cosmology obtained from the MCMC analysis, without lensing magnification (upper) and with lensing magnification (middle), and the shift (lower).
        Note that the last row in the table uses the constraints from the case with lensing magnification.
        In the top block, the analysis is run using the parametrization $\{\Omm, \Omb\}$, while the bottom block refer to the result with the parametrization $\{\omm, \omb\}$
        \label{tab:mcmc_lcdm_big_small}
    }
\begin{tabular}{r r r r r  r  r  r  r  r  r  r}
    \toprule
    \backslashbox{quantity}{$\theta$}    &    $\Omm$ & $\omm$ & $h$  & $\ns$ & $\Omb$ & $\omb$ & $\sigma_8$ \\
    \midrule
    $\sigma(\theta)$ &        0.046 & --- & 0.0425 &  0.011 & 0.0071 & --- & 0.031 \\
    $\sigma(\theta) / \theta(\%)$  &  13.031 & ---  & 6.630 &  1.152 & 13.076 & --- & 3.856 \\
    \midrule
    $\sigma(\theta)$ (L)  &        0.0425 & --- & 0.044 &  0.011 & 0.0065 & --- & 0.032\\
    $\sigma(\theta) / \theta(\%)$ (L) &  13.158 & --- &  6.567 & 1.147 & 13.206 & --- & 3.855 \\
    \midrule
    $\Delta(\theta)$ &     0.03 & ---  & \minus{0.029} & \minus{0.004} & 0.0047 & --- & \minus{0.026} \\
    $\Delta(\theta) / \sigma(\theta)$ &     0.71  & --- & \minus{0.66} & \minus{0.36} & 0.72 & --- & \minus{0.81} \\
\midrule
   \midrule
    $\sigma(\theta)$ & --- & 0.0012 & 0.043 &  0.011 & ---  & 0.00028 & 0.0315 \\
    $\sigma(\theta) / \theta(\%)$  & --- &  0.839 & 6.630 &  1.152 & --- & 1.271 & 3.856 \\
    \midrule
    $\sigma(\theta)$ (L)  & --- &        0.0425 & 0.044 &  0.011 & --- & 0.0065 & 0.032\\
    $\sigma(\theta) / \theta(\%)$ (L) & --- &  0.837 &  6.567 & 1.147 & --- & 1.274 & 3.855 \\
    \midrule
    $\Delta(\theta)$ & ---  &     0.0003  & \minus{0.029} & \minus{0.004} & ---  & 0.00005 & \minus{0.026} \\
    $\Delta(\theta) / \sigma(\theta)$ & ---  &     0.71  & \minus{0.66} & \minus{0.36} & --- & 0.18 & \minus{0.81} \\
        \bottomrule
\end{tabular}
\end{table*}

\end{document}